\documentclass[twocolumn,showpacs,english,superscriptaddress,preprintnumbers,amsmath,amssymb,floatfix]{revtex4-1}
\usepackage[T1]{fontenc}
\usepackage[latin9]{inputenc}
\usepackage{latexsym}
\usepackage{dcolumn}
\usepackage{bm}
\usepackage{graphicx}
\usepackage{color}
\usepackage{babel}
\usepackage{amsfonts}
\usepackage{enumerate}
\usepackage{mathbbol}
\usepackage{braket}

\begin{document}
\title{Quantum transport in coupled Majorana box systems}

\author{Matthias Gau}
\affiliation{Institut f\"ur Theoretische Physik, 
Heinrich-Heine-Universit\"at, D-40225 D\"usseldorf, Germany}

\author{Stephan Plugge}
\affiliation{Institut f\"ur Theoretische Physik, 
Heinrich-Heine-Universit\"at, D-40225 D\"usseldorf, Germany}

\author{Reinhold Egger}
\affiliation{Institut f\"ur Theoretische Physik, Heinrich-Heine-Universit\"at, D-40225 D\"usseldorf, Germany}

\date{\today}

\begin{abstract}
We present a theoretical analysis of low-energy quantum transport in coupled Majorana box devices. A single Majorana box represents a Coulomb-blockaded mesoscopic superconductor proximitizing 
two or more long topological nanowires. The box thus harbors at least four Majorana zero modes (MZMs).
Setups with several Majorana boxes, where MZMs on different boxes are tunnel-coupled via short nanowire
segments, are key ingredients to recent Majorana qubit and code network proposals. 
We construct and study the low-energy theory for multi-terminal junctions with normal 
leads connected to the coupled box device by lead-MZM tunnel contacts.  Transport experiments 
in such setups can test the nonlocality of Majorana-based systems and the integrity of the underlying Majorana qubits. 
For a single box, we recover the previously described topological Kondo effect which can be 
captured by a purely bosonic theory. For several coupled boxes, however,
non-conserved local fermion parities require the inclusion of additional local sets of Pauli operators. 
We present a renormalization group analysis and develop a nonperturbative strong-coupling 
approach to quantum transport in such systems.  Our findings are illustrated for several examples, including a loop qubit device and different two-box setups.  
\end{abstract}


\maketitle
\section{Introduction}

Topological superconductors harboring spatially localized Majorana bound states (MBSs)
continue to attract a lot of interest; for reviews, see  
Refs.~\cite{Alicea2012,Leijnse2012,Beenakker2013,Aguado2017,Lutchyn2018}.  
When different MBSs are located sufficiently far away from each other,
they represent fractionalized zero-energy modes:   
a pair of Majorana zero modes (MZMs) is equivalent to a single fermionic zero mode.  
Apart from the fundamental interest in experimental observations of such
exotic excitations, the potential availability of systems with robust MZMs 
holds significant promise for applications in  topological quantum information processing  \cite{Kitaev2001,Sau2010,Alicea2011,Hyart2013,Terhal2012,Sarma2015,Vijay2015,Landau2016,Plugge2016c,
Aasen2016,Vijay2016,Manousakis2017,Plugge2017,Karzig2017,Litinski2017,Litinski2018}.
It is therefore quite exciting that experiments have already provided evidence for MBSs in  
hybrid superconductor-semiconductor nanowire platforms
\cite{Mourik2012,Deng2012,Das2012,Rokhinson2012,Chang2015,Krogstrup2015,
Higginbotham2015,Albrecht2016,Deng2016,Guel2017,Albrecht2017,Nichele2017,Zhang2017,Suominen2017,
Gazi2017,Zhang2018,Deng2018,Laroche2018} 
as well as in other material classes \cite{Yazdani2014,Franke2015,Sun2016,Feldman2017,Deacon2017}.

A particularly attractive candidate for realizing a MZM-based qubit results from
mesoscopic superconducting islands containing four (or more) MZMs.  For such a floating island, 
termed Majorana box (or simply box) in what follows,
 the Coulomb charging energy $E_C$ plays a dominant role and has to be carefully taken into account \cite{Fu2010,Zazunov2011,Hutzen2012,Plugge2015}.
Under Coulomb valley conditions, the charge on the island is quantized and the box ground state conserves fermion
parity.  For a box with four MZMs, one then encounters  
a two-fold degenerate ground state which is equivalent to an effective 
spin-$1/2$ degree of freedom (qubit) nonlocally built from Majorana states \cite{Plugge2017}.
By arranging tunnel-coupled Majorana boxes in extended two-dimensional (2D) network structures, one obtains  
topologically ordered phases such as the toric code  \cite{Xu2010,Terhal2012,Nussinov2012,Landau2016,Roy2017}.
Such phases could be useful for quantum information processing applications, e.g., to implement a
Majorana surface code \cite{Terhal2012,Landau2016,Plugge2016c}.
We note that recent work has also discussed a parafermionic generalization of the Majorana box \cite{Snizkho2018}. 

On the other hand, for just a single Majorana box,
the spin-$1/2$ degree of freedom encoded by the MZMs will be subject to Kondo screening processes
if at least three normal leads are connected to the box by tunnel couplings \cite{Beri2012,Altland2013,Beri2013,Zazunov2014,Altland2014,Eriksson2014,Galpin2014,
Eriksson2014b,Kashuba2015,Buccheri2015,Plugge2016b,Herviou2016,Zazunov2017,Bao2017,
Beri2017,Landau2017,Michaeli2017}. 
Recalling that Majorana states have a well-defined spin polarization direction \cite{Aguado2017}, 
for the case of point-like tunnel contacts, the leads can be modeled as effectively 
spinless one-dimensional (1D) noninteracting electrons \cite{Alicea2012}.  
(We note that Coulomb interactions in the leads have been studied in this context \cite{Altland2013,Beri2013}, but we will not address such effects here.)
The exchange couplings of the standard Kondo problem \cite{Gogolin1998,Altland2010} are now
generated from cotunneling processes connecting different leads through the box, 
where the lead index takes over the role of the spin up/down quantum number.  
At low energy scales, such screening processes drive the system 
towards a stable non-Fermi liquid fixed point of overscreened multi-channel Kondo 
character, the topological Kondo point \cite{Beri2012}. 
From the viewpoint of multi-terminal junction theory \cite{Nayak1999,Oshikawa2006}, 
it is remarkable that this topological Kondo effect (TKE) admits a purely bosonic description via Abelian bosonization 
 for the 1D leads \cite{Altland2013,Beri2013}.  In fact, the physics is then equivalent to 
the quantum Brownian motion of a particle in a periodic 2D lattice potential
which in turn admits an exact solution at very low energies \cite{Yi1998,Yi2002}.
 
The main goal of this paper is to explore the intermediate situation between 
 just a single box connected to leads (i.e., the single-impurity TKE) and 
 an extended 2D coupled-box network.  
For instance, consider two Majorana boxes connected by tunnel links, where each box in turn 
is coupled to at least three normal leads.  Such a setup can be viewed as a 
topological Kondo variant of the celebrated two-impurity Kondo problem \cite{Jayprakash1981,Jones1988,Affleck1995}.
In the latter, one encounters a non-Fermi liquid fixed point 
not present in the single-impurity Kondo problem. In particular, the fractional quasiparticle charge for
the single-impurity topological Kondo problem, which could be probed by shot noise \cite{Zazunov2014,Beri2017} or via the Josephson effect \cite{Zazunov2017}, 
could now have a different value for the two-impurity setup.  
With predictions for transport properties of a coupled box device at hand, 
measurements of the conductance between a given pair of leads, e.g., as a function of temperature or bias voltage, can 
then yield precious insights about nonlocality effects due to MZMs.  Most importantly, by decoupling (or adding) another lead distinct from the pair of leads defining the conductance measurement, 
one expects a drastic effect on the conductance value \cite{Beri2012,Zazunov2014}.
Transport measurements could thereby establish that Majorana physics really is
behind the device functionality. 

In order to address transport and Kondo physics in coupled Majorana box devices in a 
comprehensive way, we start in Sec.~\ref{sec2} by describing a
theoretical framework suitable for tackling such problems.  In particular, we show that
Abelian bosonization \cite{Gogolin1998} in combination with the Klein-Majorana fusion 
approach of Refs.~\cite{Altland2013,Beri2013} allows for a highly versatile
 formulation of the theory.
In Sec.~\ref{sec3}, we present a detailed study of the weak-coupling regime by means of a one-loop
renormalization group (RG) analysis.  Loosely speaking, the weak-coupling regime is realized at energies 
above a suitably defined Kondo scale.  
We find that the system generally flows towards strong coupling, where in marked contrast to the
single-impurity TKE \cite{Altland2013,Beri2013}, an effectively 
bosonic description no longer applies.  In general, one has to take into account additional
non-conserved local fermion parities which can be represented by sets of Pauli operators.
Such spin-like variables are shown to play a crucial role for an understanding of transport in basically all coupled Majorana box devices. 
In Sec.~\ref{sec3}, we also provide an explicit RG analysis for three device examples of current experimental
interest, including the `loop qubit' device proposed in Ref.~\cite{Karzig2017}. 
Next, in Sec.~\ref{sec4}, we turn towards the strong-coupling regime approached at very low energy 
scales.  By focusing on the most relevant degrees of freedom,
which can be identified from the weak-coupling RG flow and by employing
quantum Brownian motion arguments \cite{Yi1998,Yi2002}, we derive and study the effective low-energy theory 
corresponding to this regime.  Employing also Emery-Kivelson-type transformations  \cite{Emery1992,Gogolin1998,Fabrizio1995,Gogolin2006,Mitchell2016,Landau2017b},
 Sec.~\ref{sec4} provides a nonperturbative strong-coupling analysis for all three examples studied 
 in Sec.~\ref{sec3} from the weak-coupling perspective. 
Finally, in Sec.~\ref{sec5}, we present the exact solution for quantum transport in a simple two-box device at a Toulouse point which exhibits two-channel Kondo physics.   
Finally, we offer some conclusions in Sec.~\ref{sec6}.
Technical details have been delegated to several Appendices, and we put $\hbar=k_B=1$ and the density of states in the leads $\nu = 1$ throughout.

In most chapters below, we include general sections introducing broadly applicable concepts and ideas of how to tackle transport in coupled Majorana boxes, followed by select simple examples that are of current interest. To follow the general discussion, the interested reader may find it useful to seek clarity about concrete applications in one or two of these examples, and to revisit the general discussion once those are understood.

\section{Model and low-energy approach} \label{sec2}

\begin{figure}
\centering  
\includegraphics[width=0.45\textwidth]{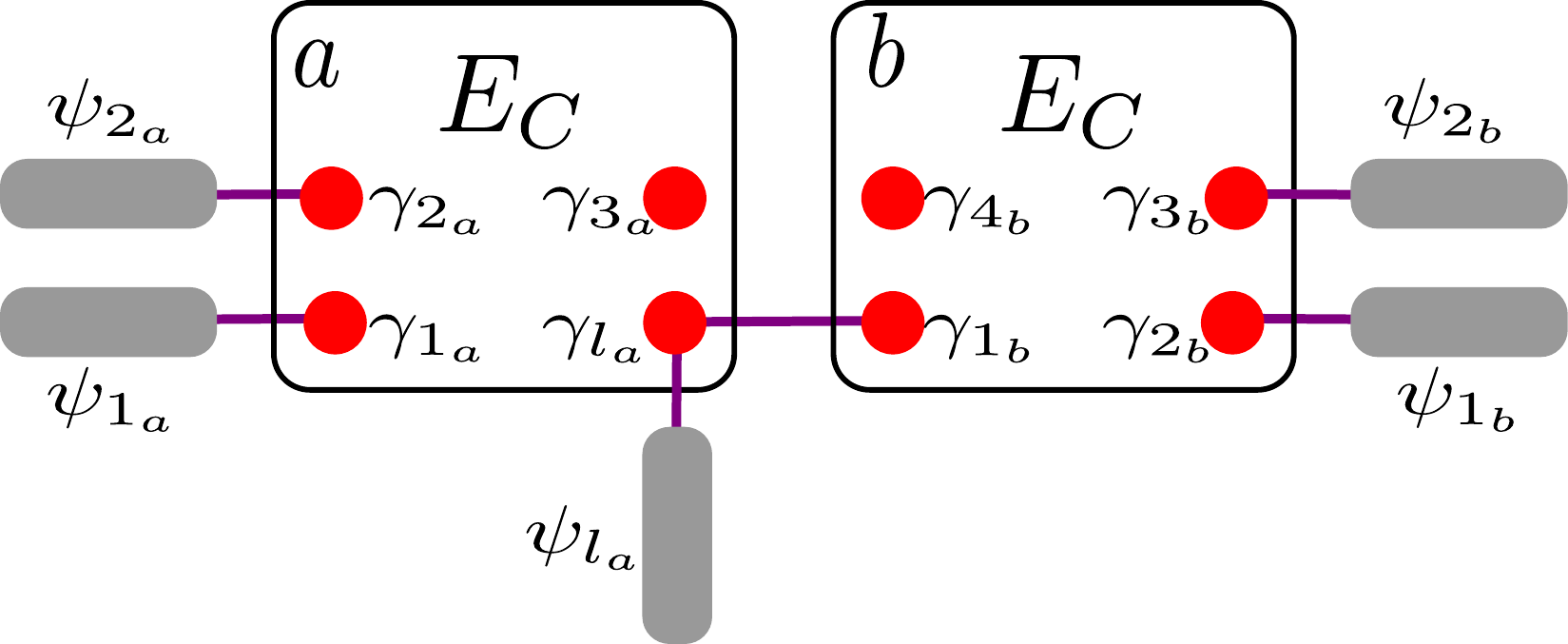}
\caption{
Example for a device with two Majorana boxes ($a,b$) connected by a single tunnel bridge (violet). 
Each box is subject to a charging energy $E_C$ and hosts four MZMs with corresponding 
Majorana operators $\gamma_{j_{a/b}}$ (filled red circles). 
Both boxes are connected to several normal leads, with corresponding fermion 
operators $\psi_{j_{a/b}}(x)$ (indicated in grey), via lead-MZM tunnel links (violet). 
For box $a/b$, we have $M_{a/b}$ simple lead-MZM tunnel contacts. Simple contacts 
are characterized by an only pairwise coupling between a lead fermion operator $\Psi_{j_a}=\psi_{j_a}(0)$ and 
a MZM operator $\gamma_{k_a}$, see Eq.~\eqref{eq:Htunlead}, without couplings to other leads or MZMs. 
For the shown case with $M_a=M_b=2$, the only non-simple contact corresponds to 
lead fermion $\psi_{l_a}$.  }
\label{fig1}
\end{figure}

The central goal of this work is to understand the low-energy physics of multi-terminal junctions defined by a set of noninteracting normal-conducting leads with point-like tunnel contacts to a general coupled Majorana box device.  A concrete example for such a setup is shown in Fig.~\ref{fig1}. 
 We start in Sec.~\ref{sec2a} by describing the basic model employed here and the physical assumptions behind it.  
 For point-like lead-MBS tunnel contacts,  it is well known that noninteracting leads can be modeled 
as effectively 1D spinless leads \cite{Gogolin1998,Altland2010,Alicea2012}. 
Subsequently,  in Sec.~\ref{sec2b} we express these 1D lead fermions in terms of Abelian 
 bosonization  \cite{Gogolin1998}, which offers a convenient route to access the important 
low-energy modes.  Tunneling processes are then analyzed in Sec.~\ref{sec2c}. Finally,
in Sec.~\ref{sec2d}, we focus on Coulomb valley conditions and describe the effective low-energy theory 
projected to the charge ground state  of each Majorana box in the system.   

\subsection{Model}\label{sec2a}

Let us start with the description of a single Majorana box, which for the moment is assumed 
decoupled from all other boxes and from all leads.
For concrete layout proposals, see Refs.~\cite{Plugge2017,Karzig2017}.
Following the discussion in Refs.~\cite{Beri2012,Altland2013,Beri2013,Zazunov2014},  
on energy scales well below the proximity-induced topological superconducting gap $\Delta$, we 
can neglect above-gap quasiparticle excitations.   
In addition, throughout this work, we will assume that all MBSs on a given box are located far away from each other and
therefore can be viewed as MZMs.  (For a discussion of hybridization effects between MBSs on a given box,
see Ref.~\cite{Altland2014}.)
Under these conditions, we only need to take into account Cooper pairs and MZMs, where Majorana operators are self-adjoint, 
 $\gamma_j = \gamma_j^\dagger$, and obey the Clifford
 algebra $\{\gamma_j,\gamma_{k}\}=2\delta_{jk}$ 
 \cite{Alicea2012,Leijnse2012,Beenakker2013,Aguado2017,Lutchyn2018}. 
We now take into account the box charging energy $E_C$, 
where $E_C\approx 1$~meV for typical experimental realizations \cite{Albrecht2016}. This 
energy scale plays a central role for all coupled box devices studied below.  
In particular, it facilitates phase-coherent electron transport, which in turn generates non-trivial
correlations between different boxes and/or leads. This basic mechanism is also behind many 
recently proposed quantum information processing schemes for Majorana qubits and Majorana
code networks \cite{Landau2016,Plugge2016c,Aasen2016,Vijay2016,
Manousakis2017,Plugge2017,Karzig2017,Litinski2017,Litinski2018}.

Under the above conditions, the Hamiltonian of an isolated box is solely due to Coulomb charging, 
\begin{equation}\label{eq:Hcharge}
H_{\rm box}=E_C\left(\hat{Q}-n_{g} \right)^2,
\end{equation}
where the dimensionless parameter $n_g$ is controlled by backgate voltages.  
We assume the same value of $E_C$ for all boxes below since 
 different charging energies do not cause qualitative changes as
 long as they remain sufficiently large.  
 The operator $\hat{Q}$ has integer eigenvalues $Q$ and describes the total charge on the box
 in units of the elementary charge $e$.  In general, $\hat Q$ receives contributions both from Cooper pairs and 
 from the MZM sector. However, it is most convenient to adopt a gauge 
 where the Majorana operators do not carry charge but instead are accompanied by 
 $e^{\pm i \varphi}$ operators whenever the box charge changes by one unit, $Q\to Q\pm 1$ \cite{Fu2010}.
By this choice, $\varphi$ is the phase operator conjugate to $\hat Q$, i.e.,
$[\varphi,\hat{Q}]=i$.
For each Majorana box, the charge dynamics is therefore captured by a dual pair of local 
bosonic fields.
For illustrative purposes, we consider boxes harboring four MZMs below. The generalization of our approach to an arbitrary even number of MZMs for a given box is straightforward.

Next we include the effects of a single MZM-MZM tunnel link connecting 
two Majorana boxes $a/b$, cf.~Fig.~\ref{fig1}, via the tunneling Hamiltonian \cite{Fu2010,Zazunov2011}
\begin{equation}\label{eq:Htunisland}
H_t = t_{j_ak_b}\gamma^{}_{j_a}\gamma^{}_{k_b} e^{i(\varphi_a-\varphi_b)} +\mathrm{h.c.}
\end{equation}
with the MZM operators 
$\gamma_{j_a}$ and $\gamma_{k_b}$.  The index $j_a$ ($k_b$) here means that we label MZMs belonging to
 box $a$ ($b$), cf.~Fig.~\ref{fig1}, and 
 the $e^{\pm i\varphi_{a,b}}$ operators describe the transfer of charge in a tunneling event. 
 Physically, the $e^{i(\varphi_a-\varphi_b)}$ factor in Eq.~\eqref{eq:Htunisland} 
amounts to the formation of a charge dipole between both boxes. 
Finally, $t_{j_a k_b}$ is a microscopic tunnel amplitude connecting the respective MZMs, 
e.g., through an intermediate non-topological nanowire segment. 

For point-like lead-MZM tunnel contacts, we can now describe each 
noninteracting lead by a 1D spinless fermion operator 
$\psi_{j_a,R/L}(x)$ \cite{Gogolin1998,Altland2010,Alicea2012}, 
where the index $j_a$ indicates that the lead is tunnel-coupled to box $a$. 
Choosing $x=0$ as the tunnel-contact point, 
right- and left-moving ($R/L$) fermions are defined for $x<0$, with  
the open boundary conditions $\psi_{j_a,L}(0)=\psi_{j_a,R}(0)$.
By a standard unfolding transformation \cite{Gogolin1998}, we may switch to  chiral (right-moving)
fermions, $\psi_{j_a}(x)$, by writing $\psi_{j_a}(x)=\psi_{j_a,R}(x)$  
for $x<0$ and $\psi_{j_a}(x)=\psi_{j_a,L}(-x)$ for $x>0$. 
The lead-MZM contact is then described by the tunneling Hamiltonian 
\begin{equation}\label{eq:Htunlead}
H_\lambda = \lambda_{j_ak_a}\Psi_{j_a}^\dagger\gamma_{k_a}^{} e^{-i\varphi_a} +\mathrm{h.c.},
\end{equation}
where $\lambda_{j_ak_a}$ again is a microscopic tunneling amplitude and
 we employ the shorthand notation $\Psi_{j_a}= \psi_{j_a}(0)$.
 
 All tunnel couplings will be assumed so weak that they can neither 
 create above-gap quasiparticle excitations nor destroy the integrity of MBSs.
 We thus require that the energy scales associated with the amplitudes $t_{j_ak_b}$ and $\lambda_{j_ak_a}$
  are small compared to both $\Delta$ and $E_C$.
Moreover, we note that physical tunnel contacts  
extend only over short distances within the coupled box device.
The only exception to this rule are long-ranged pairwise cotunneling events generated via charging effects, see Sec.~\ref{sec2d} below.

Finally, the Hamiltonian of decoupled lead no.~$j$  is given by 
\begin{equation}\label{eq:Hleadsfermion}
H_\mathrm{leads} = -iv_F \int_{-\infty}^\infty dx \ \psi^\dagger_{j}\partial_x\psi_{j},
\end{equation}
where we assume the same Fermi velocity $v_F$ for all leads and write $j=j_a$ for notational simplicity.  
Differences in Fermi velocities are not important and can be taken into account by  renormalizing the above 
tunneling amplitudes.  

\subsection{Abelian bosonization}\label{sec2b}

So far we have considered a fermionic description of the leads. By inspecting the tunneling 
Hamiltonians \eqref{eq:Htunisland} and \eqref{eq:Htunlead}, 
we observe that it will also be useful to switch to a bosonized description for the leads.
As for the Majorana box above, fermionic (statistical) and bosonic (charge/phase) lead variables are
thereby explicitly separated. While the lead Hamiltonian \eqref{eq:Hleadsfermion} 
admits a purely bosonic description, see Eq.~\eqref{eq:Hleads} below, 
fermionic aspects do appear in tunneling operators connecting the respective lead to MZMs or to other leads.
In terms of right- and left-movers, Abelian bosonization states the correspondence \cite{Gogolin1998}
\begin{equation}\label{eq:boso}
\psi_{j,R/L}^{\dagger}(x)= \frac{\kappa_j}{\sqrt{\alpha}} e^{i[\phi_{j}(x)\pm\theta_{j}(x)]}
\end{equation}
with a short-distance cutoff length $\alpha$. 
The dual boson fields $\phi_{j}$ and $\theta_{j}$ obey the algebra 
 $[\phi_{j}(x'),\partial_x\theta_{k}(x)]=i\pi\delta\left(x-x'\right)\delta_{jk}$, and $\kappa_{j}$ 
 denotes a Klein factor ensuring anticommutation relations with all other lead fermions and all MZM operators.   
Following Refs.~\cite{Altland2013,Beri2013}, we use a Majorana fermion representation for 
Klein factors, i.e.,  
$\kappa_{j}^{\dagger}=\kappa_{j}$ and $\lbrace\kappa_{j},\kappa_{k}\rbrace=2\delta_{jk}$.  
Noting that the open boundary conditions for lead fermions 
translate to $\theta_{j}(0)=0$, and using the shorthand notation
\begin{equation}\label{shorthand}
\Phi_{j}= \phi_j(0),\quad \Theta_{j}'  = \partial_x\theta_{j}(0),
\end{equation}
the lead fermion operator in Eq.~\eqref{eq:Htunlead} takes the form
$\Psi^\dagger_{j}=\alpha^{-1/2}\kappa_{j} e^{i\Phi_{j}}$.
Similarly, the electron density operator near the tunnel contact is proportional to 
$\Theta_{j}'$. 

The lead Hamiltonian \eqref{eq:Hleadsfermion} is given by \cite{Gogolin1998}
\begin{equation}\label{eq:Hleads}
H_\mathrm{leads}=\frac{v_F}{2\pi}\int_{-\infty}^0 dx\left[\left(\partial_x\phi_{j}\right)^2+
\left(\partial_x\theta_{j}\right)^2 \right].
\end{equation}
For a description of tunneling processes, however, Klein factors play a crucial role. Using bosonized expressions,
each tunneling event is factorized into a charge-neutral fermion-bilinear part encoding 
the fermionic statistics and a part describing the bosonic charge (or phase) dynamics. 
Explicitly, for the lead-MZM tunneling Hamiltonian in Eq.~\eqref{eq:Htunlead}, we obtain
\begin{equation}\label{eq:Htunlead2}
H_\lambda  = \lambda_{j_ak_a} \kappa_{j_a}\gamma_{k_a}
e^{i(\Phi_{j_a}-\varphi_a)}+\mathrm{h.c.},
\end{equation}
where a factor $1/\sqrt{\alpha}$ has been absorbed in
 $\lambda_{j_a k_a}$.  We notice that Eq.~\eqref{eq:Htunlead2} contains a local 
 fermion parity operator $i\kappa_{j_a}\gamma_{k_a}$ with eigenvalues $\pm 1$
  corresponding to the occupation number of the 
 fermion mode built from $\kappa_{j_a}$ and $\gamma_{k_a}$. 
 
 \subsection{Simple vs non-simple contacts}
\label{sec2c}

 \begin{figure}
\centering  
\includegraphics[width=0.4\textwidth]{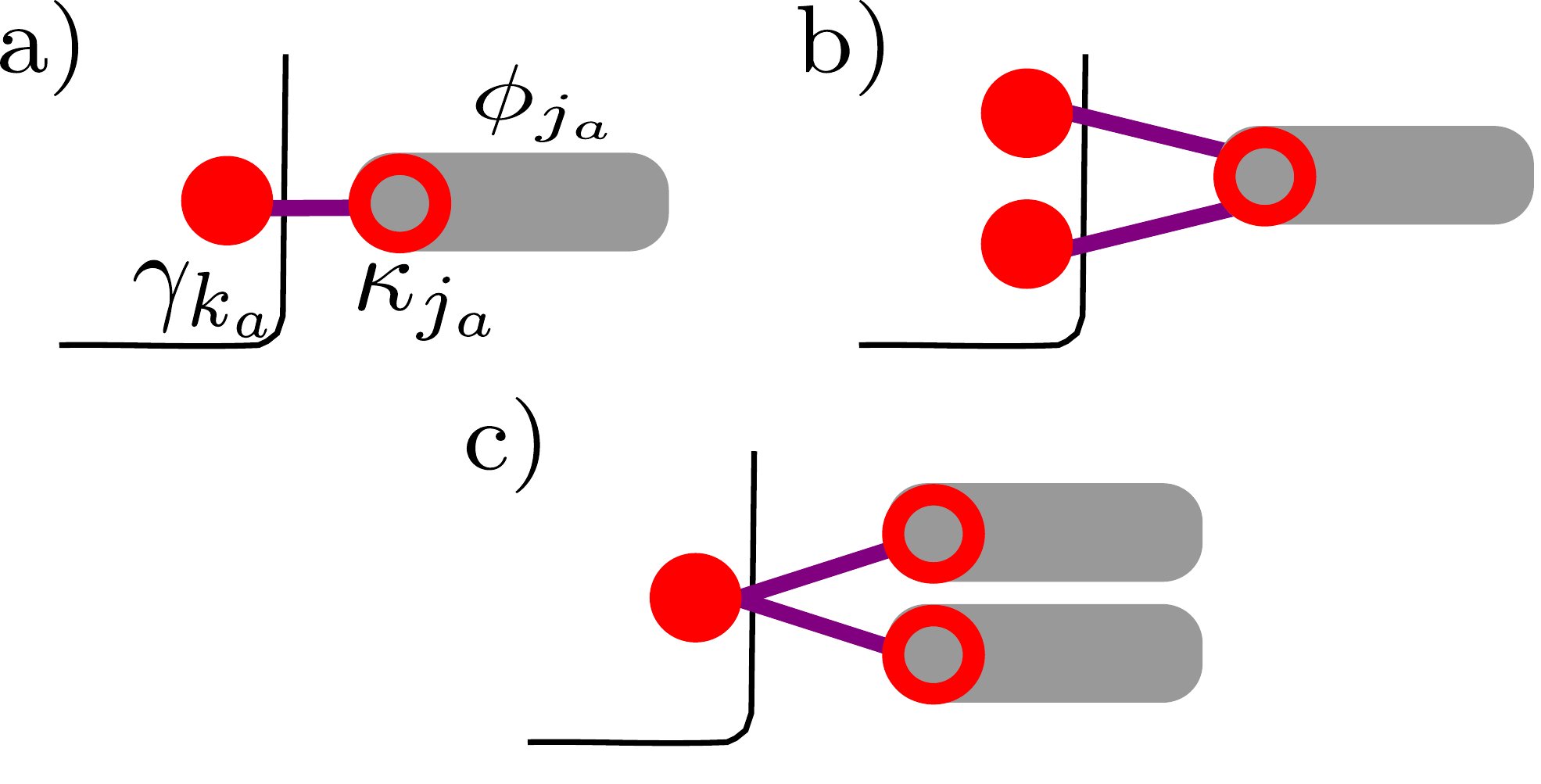}
\caption{Simple vs non-simple lead-MZM tunnel junctions, see Sec.~\ref{sec2c}. Filled red circles 
correspond to MZMs $\gamma_{k_a}$ and open red circles to Klein-Majorana operators $\kappa_{j_a}$
within a bosonized description of lead fermions, see Eq.~\eqref{eq:boso}.
(a) Simple contact, cf.~Eq.~\eqref{eq:Htunlead2}.  
(b) Non-simple contact between two MZMs and one lead, cf.~Eq.~\eqref{eq:Htun21}.
(c) Non-simple contact between one MZM and two leads, cf.~Eq.~\eqref{eq:Htun12}.}\label{fig2}
\end{figure}
 
It is convenient for the subsequent discussion to introduce the notion of a \emph{simple} lead-MZM contact,
and generally that of a simple tunnel junction.
For a simple contact, see Fig.~\ref{fig2}(a), we require that the tunnel-coupled
Majorana ($\gamma_{k_a}$) and lead ($\Psi_{j_a}$) fermions have no additional tunnel couplings 
to other (Majorana or lead) fermions. 
All lead-MZM junctions beyond the pairwise tunnel contact in Fig.~\ref{fig2}(a) are 
referred to as \emph{non-simple}.
Two examples of such non-simple lead-MZM contacts are shown in Figs.~\ref{fig2}(b) and (c),
see also Ref.~\cite{Affleck2013}.
A non-simple junction also occurs when a lead-contacted MZM is in addition tunnel-coupled to another MZM on an adjacent box, see Fig.~\ref{fig1}.
Similarly one may refer to non-simple MZM-MZM junctions if several MZMs on distinct boxes are coupled to each other.

For systems with only simple contacts, we
can then proceed in a straightforward manner by employing the Klein-Majorana fusion approach
put forward in Refs.~\cite{Altland2013,Beri2013}.
To that end, we observe that in such systems, each local fermion parity
built from a Klein-Majorana operator $\kappa_{j_a}$ and a MZM 
operator $\gamma_{k_a}$ forming the respective tunnel contact, cf.~Fig.~\ref{fig2}(a), 
will be separately conserved, $i\kappa_{j_a}\gamma_{k_a}=\pm 1$.
Similarly, all local parities associated with MZM-MZM tunnel links 
are conserved, $i\gamma_{j_a}\gamma_{k_b}=\pm 1$. 
The above observations imply that the fermionic sector of the theory is trivially solvable so long as 
 all local fermion parities remain conserved. 
 A coupled Majorana box system with only simple contacts can thus be reduced 
 to a purely bosonic theory, which is generally much simpler to analyze than the original
 fermionic version.

In this work, we address situations where some of the above 
local fermion parities are not conserved anymore.
This may happen if unintentional parity-breaking mechanisms are present, e.g., when a
conventional mid-gap Andreev state is accidentally centered near a lead-contacted MBS and thereby activates quasi-particle poisoning mechanisms \cite{Plugge2016b}. We instead will 
focus on intentional parity-breaking effects due to non-simple tunnel contacts.
Such cases pertain to many Majorana box transport setups and
quantum-information processing applications. 
In fact, local parity conservation implies that for systems with only simple contacts,  MZMs
cannot reveal their underlying fermionic statistics since different measurement bases 
are not accessible. With the above motivation, we now inspect several generic scenarios.

\subsubsection{Charge degenerate boxes}

Our first example for parity-breaking mechanisms
is tied to fluctuating charge states on a given box, e.g., because the gate 
parameter $n_g$ in Eq.~\eqref{eq:Hcharge} is tuned close to a half-integer value. 
This case has also been studied in the context of the single-impurity TKE \cite{Herviou2016,Michaeli2017,Landau2017}. In general, a large box charging energy $E_C$ will 
admit at most a few low-energy charge states.
As a consequence, charging effects also constrain the box fermion parity which can be written as 
the product of MZM operators on the box.
For the four-MZM box \cite{Beri2012}, we have ${\cal P}_{\rm box}=\gamma_1\gamma_2\gamma_3\gamma_4$. 

For $n_g$ close to an half-integer value and/or for strong lead-MZM tunnel couplings, 
the box charge can fluctuate strongly.
Retaining only the nearly degenerate lowest-energy  charge states 
$\ket{Q}_a$ and $\ket{Q+1}_a$ on box no.~$a$, where the integer $Q$ is 
chosen such that $Q<n_g<Q+1$, it is convenient to introduce a corresponding spin-$1/2$  operator ${\bf S}_a$.
With $S_{\pm,a}=S_{x,a}\pm i S_{y,a}$, it has the components \cite{Fu2010}
\begin{eqnarray}
  S_{z,a} &=&\left(\ket{ Q+1}_a-\ket{ Q}_a \right)/2, \\ 
   S_{+,a}&=& S_{-,a}^\dagger =\nonumber  e^{ i\varphi_a}= \ket{Q+1}_a\bra{Q}_a.
  \end{eqnarray} 
Projecting Eqs.~\eqref{eq:Hcharge}, \eqref{eq:Htunisland} and \eqref{eq:Htunlead2} to the 
Hilbert subspace spanned by $\ket{Q}_a$ and $\ket{Q+1}_a$, 
 the Hamiltonian schematically takes the form \cite{Herviou2016,Michaeli2017}
\begin{eqnarray}
H_\mathrm{deg}
&=& \Delta E_a S_{z,a}+\sum_{j_a, k_b} \left( 
t_{j_a k_b}\gamma_{j_a} \gamma_{k_b} S_{+,a}e^{-i\varphi_b} +{\rm h.c.}\right)
\nonumber \\
&+& \sum_{j_a,k_a}\left( \lambda_{j_a k_a}\gamma_{j_{a}}\kappa_{k_a} 
S_{+,a} e^{-i\Phi_{k_a}}
+\mathrm{h.c.}\right),\label{eq:Hchargespin}
\end{eqnarray}
where the energy $\Delta E_a$ is controlled by the detuning of $n_g$ away from half-integers
and we use the definition in Eq.~\eqref{shorthand}.
While $H_\mathrm{deg}$ in Eq.~\eqref{eq:Hchargespin} allows ${\cal P}_{\rm box}$ to fluctuate, 
such fluctuations are perfectly correlated with charge hopping processes on and off the box:
the MZM operator $\gamma_{j_a}$ is always accompanied by  $S_{\pm,a}$, see 
Eq.~\eqref{eq:Hchargespin}.
As long as the system only has simple lead-MZM contacts, one therefore arrives at 
a purely bosonic description again. In fact, while details of the single-impurity TKE  
such as the value of the Kondo temperature depend on the backgate parameter, 
the low-energy behavior is basically independent of $n_g$ \cite{Herviou2016,Michaeli2017}.
By implementing an entangled lead-MZM fermion basis, the Klein-Majorana fusion approach
 is thus highly useful also for charge-degenerate Majorana box devices.  
 We will see that this conclusion applies even in a much wider sense. 

\subsubsection{Non-simple contacts}

Next we consider device layouts with at least one non-simple contact
where in- or out-tunneling of charge from the box can take place either
 via different MZMs on the box [Fig.~\ref{fig2}(b)] or
  through different leads [Fig.~\ref{fig2}(c)].
The presence of such contacts has important consequences on low-energy quantum transport in 
coupled Majorana box junctions since the corresponding local fermion parities defined above are not conserved
anymore.  In particular, after a sequence of tunneling events, some of these parities may have been flipped  
along with a charge transfer between different leads. Similar processes 
have been discussed in Refs.~\cite{Kashuba2015,Plugge2016b} and are known to affect  
transport properties.

To make progress, it is useful to identify subsets of (MZM and Klein factor) Majorana operators with 
conserved overall parity. Such a subset must contain an even number $m$ of 
Majorana operators, where the corresponding Majorana bilinears  generate a spin operator 
with symmetry group SO$(m)$ \cite{Beri2012,Altland2013,Beri2013,Altland2014}.
For both cases in Figs.~\ref{fig2}(b,c),  three Majorana operators are coupled together
at the junction. Taking into account a dummy Majorana mode not shown in Fig.~\ref{fig2},
the parity associated with these Majorana states is conserved.  
As a consequence, the Majorana bilinears resulting from this subset 
can equivalently be described by Pauli operators $\sigma_{x,y,z}$ as we discuss next.

As first example, consider the situation in Fig.~\ref{fig2}(b), where two Majorana operators 
($\gamma_x,\gamma_y$) on the same box (with phase $\varphi$ conjugate to $\hat Q$) 
 are tunnel-coupled with amplitudes $\lambda_{x,y}$
 to a single lead. The latter is described by the fermion operator $\Psi^\dagger \sim\kappa e^{i\Phi}$. 
 Including for completeness also a finite overlap integral between  the MBSs ($h_z$),
the tunneling Hamiltonian \eqref{eq:Htunlead} for such a junction takes the form
\begin{eqnarray}\label{eq:Htun21}
H_{2,1} &=& (\lambda_x\sigma_x+\lambda_y\sigma_y)e^{i(\Phi-\varphi)}+\mathrm{h.c.} + h_z \sigma_z
,\\  \nonumber
&& \quad \sigma_{x,y} =i\kappa\gamma_{x,y} ,\quad \sigma_z = i\gamma_y\gamma_x.
\end{eqnarray}
For a specific phase relation between $\lambda_x$ and $\lambda_y$, the same model 
describes quasiparticle poisoning effects for the single-impurity TKE \cite{Plugge2016b}.
As shown in Ref.~\cite{Plugge2016b},
in the presence of additional leads, the RG flow will generate  
an additional hybridization term $\sim \sigma_z\Theta'$ between a Pauli operator and the 
boundary fermion density.
In Sec.~\ref{sec3d}, we will discuss how this finding generalizes to arbitrary complex $\lambda_{x,y}$.

Next we turn to the alternative setup shown in Fig.~\ref{fig2}(c), where one MZM  ($\gamma$) 
is tunnel-coupled to two leads with amplitudes $\lambda_{x,y}$, cf.~Ref.~\cite{Affleck2013} for the corresponding
$E_C=0$ case.
The respective lead fermions are now written as $\Psi^\dagger_{x,y}\sim \kappa_{x,y}e^{i\Phi_{x,y}}$.
From Eq.~\eqref{eq:Htunlead}, the tunneling Hamiltonian is then given by 
\begin{equation}
H_{1,2} = \left(\lambda_x\sigma_x e^{i\Phi_x}+\lambda_y\sigma_y e^{i\Phi_y}\right)
e^{-i\varphi}+\mathrm{h.c.},\label{eq:Htun12}
\end{equation}
where  $\sigma_{x,y} = i\gamma\kappa_{x,y}$. Note that there is no $h_z\sigma_z$ contribution 
with $\sigma_z=i\kappa_y\kappa_x$. Direct lead-lead tunneling processes (if present) would produce 
different terms. 

We also observe that as long as an arbitrary coupled box system does not admit tunneling paths forming closed loops, 
all relative phases between tunneling amplitudes can be absorbed by suitable shifts of lead 
boson fields and thus do not affect the physics.
Here closed loop configurations in Hilbert space may arise from ring exchange processes involving several boxes,
for instance, a plaquette operator in Majorana code networks
\cite{Landau2016}.  A closed loop is also found for a lead coupled to several MZMs on the same box, 
see~Fig.~\ref{fig2}(b).
As a consequence, while the relative phase between $\lambda_{x}$ and $\lambda_y$ can be gauged away 
for the case shown in Fig.~\ref{fig2}(c), this is not possible for the setup in Fig.~\ref{fig2}(b) anymore.

\begin{figure}[t]
\centering  
\includegraphics[width=0.45\textwidth]{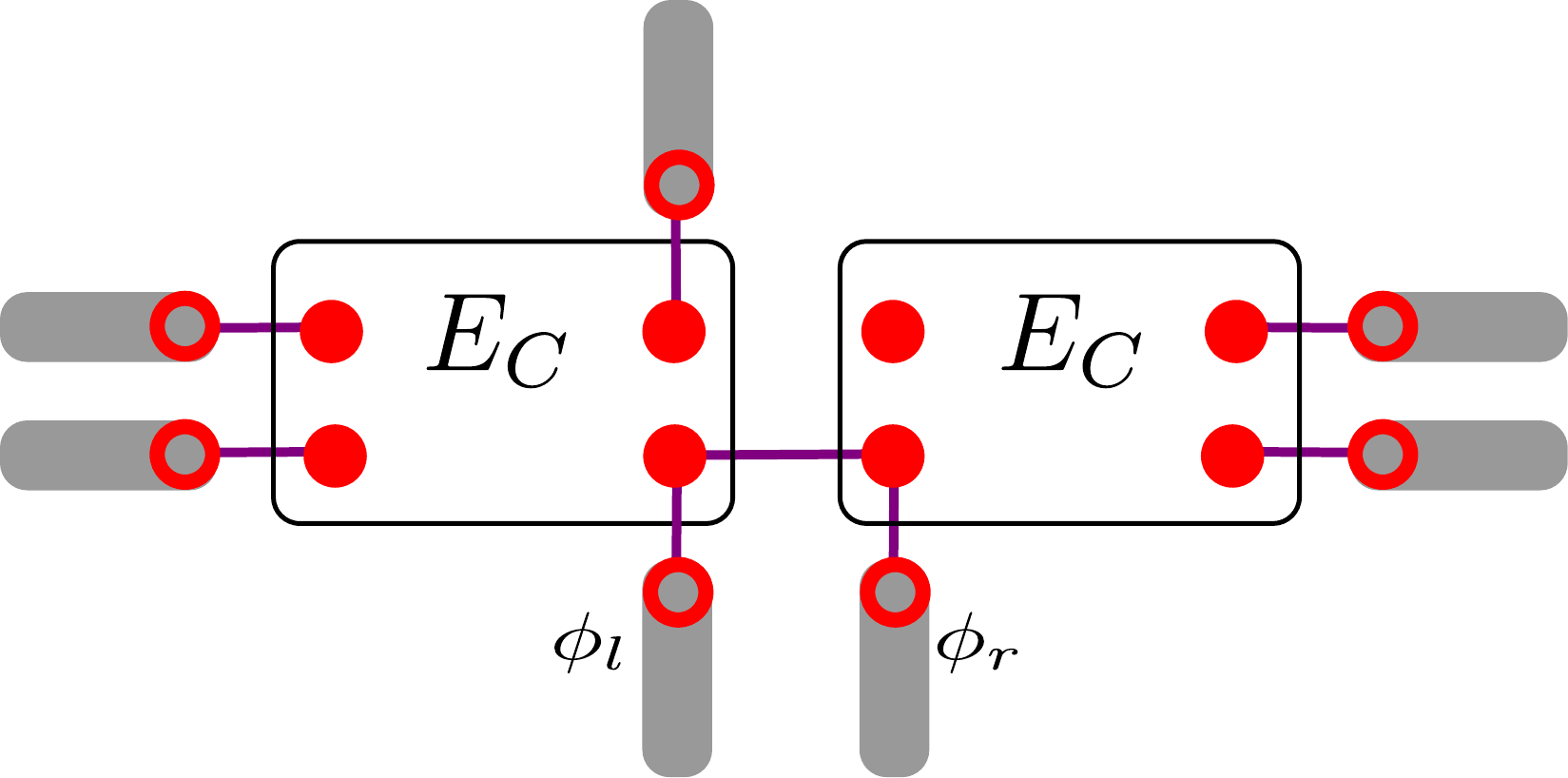}
\caption{Two-box setup with a single tunnel bridge connecting the two boxes. 
Two central leads with boson fields $\Phi_{l,r}=\phi_{l,r}(0)$ are tunnel-coupled to the respective 
MZMs, see Eq.~\eqref{eq:Htuntwobox}.
Because of the presence of the MZM-MZM link, those lead-MZM contacts are non-simple. 
In addition, $M_{L/R}$ leads are attached to the left/right box via
simple contacts, where the shown example is for $M_L = 3$ and $M_R = 2$. For an 
explanation of symbols, see Figs.~\ref{fig1} and \ref{fig2}.}
\label{fig3}
\end{figure}

As a more complicated example for a system with non-simple contacts, we next consider the two-box setup in Fig.~\ref{fig3}.
Similar setups arise in basic Majorana qubit and multi-box measurements \cite{Plugge2017,Karzig2017} 
and in the context of stabilizer codes \cite{Landau2016,Plugge2016c}.  
Here the left/right ($a=L/R$) box is connected to an arbitrary number $M_{L/R}$ of normal leads 
via simple lead-MZM contacts. Figure~\ref{fig3} shows the case $M_L=3$ and $M_R=2$. 
In addition, two central leads with the respective fermion operator $\Psi^\dagger_{l/r}\sim \kappa_{l/r}e^{i\Phi_{l/r}}$
 are connected to the left/right box through non-simple contacts to the respective MZM operator $\gamma_{l/r}$
(with tunneling amplitude $\lambda_{l/r}$). The contacts are non-simple because   $\gamma_l$ and $\gamma_r$ are tunnel-coupled by an amplitude $t_{LR}$.  
With the box phase operators $\varphi_{L/R}$, the corresponding central part of the coupled device 
is described by the Hamiltonian 
\begin{eqnarray}\label{eq:Htuntwobox}
H_{c} &=& t_{LR}\sigma_z e^{i(\varphi_L-\varphi_R)}
+\lambda_l\sigma_x e^{i(\varphi_L-\Phi_l)}\\ \nonumber
&+&\lambda_r\sigma_x e^{i(\varphi_R-\Phi_r)}+\mathrm{h.c.},
\end{eqnarray}
where we define $\sigma_z = i\gamma_l\gamma_r$ and $\sigma_x = i\gamma_l\kappa_l$.
Note that we can also write $\sigma_x\sim i\gamma_r\kappa_r$ since
the central junction parity $\gamma_l\gamma_r\kappa_l\kappa_r =\pm 1$ is conserved. 
The appearance of different Pauli operators in Eq.~\eqref{eq:Htuntwobox} suggests that for $\lambda_{l/r}\neq 0$, 
the two-box setup in Fig.~\ref{fig3} is more difficult to analyze than a purely bosonic counterpart 
with only simple contacts, e.g., without the central leads in Fig.~\ref{fig3}.

\subsection{Quantized box charge and cotunneling operators}\label{sec2d}

Our subsequent discussion will mainly focus on systems where all Majorana boxes are 
operated at near-integer $n_g$, i.e., the charge on each box has a 
quantized ground-state value.  As discussed in Sec.~\ref{sec2c}, while near-degenerate box charge states (with $n_g$ close to half-integer values) can change details of  the TKE \cite{Herviou2016,Michaeli2017}, they do not involve additional 
non-conserved fermion parity degrees of freedom (here represented by Pauli operators). 
For weak tunneling amplitudes (cf.~Sec.~\ref{sec2a}) and nearly integer $n_g$ on all boxes,
 the system is described by cotunneling amplitudes connecting in principle any pair of leads 
in the system via phase-coherent second- or higher-order charge tunneling processes.
To obtain the corresponding cotunneling amplitudes in a systematic way, we have employed a Schrieffer-Wolff transformation to 
project the full theory to the quantized charge ground-state sector of all boxes, see also Refs.~\cite{Landau2016,Plugge2016c}. 

The projected cotunneling Hamiltonian will now contain qualitatively different terms.  First, there are \emph{purely bosonic}
cotunneling contributions. Such processes do not involve Pauli operators representing non-conserved fermion parities and 
 have the schematic form  
\begin{eqnarray}\label{eq:Htunboso}
H_{\rm bos} &=& J_{j_a k_b} e^{i(\Phi_{j_a}-\Phi_{k_b})}+\mathrm{h.c.},\\ \nonumber
J_{j_ak_b}&\simeq& \frac{\lambda_{j_aj_a'}\lambda_{k_bk_b'}^\ast}{E_C}
\prod_{\langle l,l'\rangle} \frac{t_{ll'}}{E_C}.
\end{eqnarray}
The cotunneling amplitude $J_{j_a k_b}$ contains the initial and final lead-MZM couplings 
$\lambda_{j_aj_a'}$ and $\lambda_{k_bk_b'}^\ast$ for charge tunneling to/from lead $j_a/k_b$ via box $a/b$, 
see Eqs.~\eqref{eq:Htunlead} and \eqref{eq:Htunlead2}. (Here, $a=b$ is possible.)
As a result of the projection to the charge ground-state sector, the  $e^{\pm  i\varphi_{a/b}}$ terms are not present 
anymore in Eq.~\eqref{eq:Htunboso} and become effectively 
replaced by $1/E_C$ factors in the cotunneling amplitude, see Ref.~\cite{Plugge2016c}.  
In order to obtain a contribution for lead pairs attached to different boxes ($a\ne b$), 
a sequence of intermediate MZM-MZM tunneling events with
respective amplitudes $t_{ll'}$, cf.~Eq.~\eqref{eq:Htunisland}, is necessary.  In order to contribute to Eq.~\eqref{eq:Htunboso}, however, such MZM-MZM links 
must have conserved local parities.  We also note that since
for each additional tunneling event, the contribution to $J_{j_ak_b}$ gets suppressed by a factor $|t_{ll'}|/E_C\ll 1$, 
the shortest tunneling path(s) between a chosen pair of leads  will dominate.

Next, in contrast to the purely bosonic case in Eq.~\eqref{eq:Htunboso}, we consider what happens if
the tunneling path connecting leads $j_a$ and $k_b$ involves a string of Pauli operators
$\sigma^m=\sigma^m_{x,y,z}$. Here $\sigma^m$ describes 
the non-conserved local fermion parity at the $m$th non-simple link along the path.     
For a string of $n\ge 1$ Pauli operators ($m=1,\ldots,n$), 
 the projected Hamiltonian has the schematic form
\begin{equation}\label{eq:Htunboso2}
H_{\rm nbos}= J_{j_ak_b}^{(\sigma^1,\ldots,\sigma^n)}\sigma^1\cdots\sigma^n
e^{i(\Phi_{j_a}-\Phi_{k_b})}+\mathrm{h.c.},
\end{equation}
where $J^{(\{\sigma\})}_{j_ak_b}$ is a cotunneling amplitude as in Eq.~\eqref{eq:Htunboso} and the superscript
serves to remind us that this amplitude applies to a specific tunneling path involving the corresponding Pauli operator string.
Concrete examples for this notation will be given in Sec.~\ref{sec3} and in App.~\ref{appA}.
We note that with the conventions 
$J_{j_a k_b}^{(\{\sigma\})}\to J_{j_a k_b}$ and $\sigma^1\cdots\sigma^n\to 1$ for $n=0$,
i.e., in absence of non-simple links, Eq.~\eqref{eq:Htunboso} constitutes just a special case of Eq.~\eqref{eq:Htunboso2}.

We close this section by addressing additional complexities in tunneling at a non-simple junction that comprises multiple Pauli operators of the same set $\sigma_{x,y,z}$.
For example, at non-simple contacts in Fig.~\ref{fig2}(b,c), elemental tunneling events may involve anticommuting Pauli operators $\sigma_x$ and $\sigma_y$.
The corresponding path contribution now exhibits an extra suppression factor $\sim|\Delta n_g|$, where $\Delta n_g$ is the detuning of the backgate parameter $n_g$ away from integer values.
This suppression arises from the destructive interference between tunneling events with different time ordering \cite{Landau2016,Plugge2016c}.
In particular, if the box is tuned precisely to a Coulomb valley center, $\Delta n_g=0$, such paths give no contribution at all.
For finite $\Delta n_g$, both Pauli operators effectively combine to the third Pauli operator, e.g., $\sigma_x\sigma_y=i\sigma_z$.
With this change and including the $|\Delta n_g|$ factor, the cotunneling contribution is then again given by Eq.~\eqref{eq:Htunboso2}.

Further, in coupled box devices allowing for closed loops, see Sec.~\ref{sec2c} and Fig.~\ref{fig2}(b), elemental tunneling events that connect to distinct MZMs may lead to the same charge transfer.
Therefore several distinct paths with different Pauli operator content can contribute to a given cotunneling term $\sim e^{i(\Phi_{j_a}-\Phi_{k_b})}$.
Such effects have been exploited, for instance, for Majorana box qubit readout and manipulation schemes  \cite{Landau2016,Plugge2016c,Plugge2017,Karzig2017}.
Below we do not consider cases with interfering paths, or if present, as for the loop qubit device in Sec.~\ref{sec3d} and \ref{sec4e}, we explicitly separate them.

\section{Renormalization group analysis}\label{sec3}

Using the composition rules for cotunneling Hamiltonians in Sec.~\ref{sec2d}, we next turn to the derivation and
analysis of the one-loop RG equations.  We study general coupled Majorana box devices under Coulomb valley
conditions, where non-conserved local fermion parities are 
described by Pauli operators $\sigma^m=\sigma^m_{x,y,z}$ at the $m$th link.
In Sec.~\ref{sec3a}, we explain how RG equations
for systems of this type can be constructed by using the standard operator product expansion (OPE) technique 
 \cite{Gogolin1998,Altland2010}.  Subsequently we will study these equations for three device examples in order 
 to illustrate typical effects caused by non-conserved local fermion parities.

\subsection{RG equations: Construction principles}\label{sec3a}

Let us consider the perturbative expansion of the partition function in powers
of the cotunneling contributions to the Hamiltonian $H$, see~Eq.~\eqref{eq:Htunboso2}.
The RG approach \cite{Altland2010} studies how cotunneling amplitudes are renormalized, and whether 
new couplings are generated, upon reducing the effective lead bandwidth $D$ from its initial value, 
$D(\ell=0)\simeq {\rm min}\{E_C,\Delta\}$.  Writing $D(\ell)=D(0)e^{-\ell}$,
the RG equations describe the physics on lower and lower energy scales with 
increasing flow parameter $\ell$.  
We show below that always at least a few cotunneling amplitudes will flow towards strong coupling.
Since perturbation theory then breaks down at sufficiently low energy scales, the RG approach 
can only describe the weak-coupling regime.   
The physics in the strong-coupling regime will be addressed in Secs.~\ref{sec4} and \ref{sec5}.

In order to obtain RG equations via the OPE approach, one considers arbitrary pairs of cotunneling operators contributing
to $H$.  For two operators acting at almost coinciding (imaginary) times $\tau$ and $\tau'$, the result of such a contraction
must be equivalent to a linear combination of all possible operators at time $(\tau+\tau')/2$, where
the respective expansion coefficients directly determine the one-loop RG equations \cite{Gogolin1998,Altland2010}.  We thus have to analyze
contractions of cotunneling operator pairs. Denoting the corresponding amplitudes by
 $J_{j m}^{(\{\sigma\})}$
and $J_{m k}^{(\{\sigma^\prime\})}$,  their contraction renormalizes the tunneling amplitude 
$J_{jk}^{(\{\sigma^{\prime\prime}\})}$, where the Pauli string $\{\sigma^{\prime\prime}\}$ follows
by multiplication of both operator strings. This composite tunneling amplitude thus connects 
leads $j$ and $k$ by a tunneling path touching lead $m$ and back.   
The RG equations now depend on whether the Pauli strings
$\sigma^1\cdots\sigma^n$ and $\sigma^{1'}\cdots\sigma^{n'}$ 
commute or anticommute.

\subsubsection{Commuting Pauli strings}

For commuting Pauli strings, the OPE approach yields the general RG equations
(lead density of states $\nu = 1$)
\begin{equation}\label{eq:RGgeneral}
\frac{dJ_{jk}^{(\{\sigma^{\prime\prime}\})}}{d\ell} = 
\sum_{m\neq (j,k)} J_{j m}^{(\{\sigma\})}J_{mk}^{(\sigma^\prime)}.
\end{equation}
This result is simple to understand if both Pauli strings do not share overlapping Pauli operators at all. 
The composite tunneling path is then obtained by simply stitching together both paths, and the Pauli string
 $\{\sigma^{\prime\prime}\}$ corresponds to the product of the strings $\{\sigma\}$ and $\{\sigma^\prime\}$.
Moreover, if identical Pauli operators appear in both strings, say, $\sigma^{m}_x$ and $\sigma^{m'=m}_x$, they 
 effectively square to unity and thus drop out in the string $\{\sigma^{\prime \prime}\}$.  Let us 
 now discuss Eq.~\eqref{eq:RGgeneral} in more detail for different cases of interest.

To that end, it is very convenient to introduce the concept of \emph{bosonic subsectors} (or simply subsectors).
A bosonic subsector ${\cal B}$ refers to a group of $M$ leads (with index $j\in {\cal B}$) which are coupled to each other through purely bosonic cotunneling processes, and hence undergo purely bosonic interactions within the subsector, cf.~Eq.~\eqref{eq:Htunboso}.
For example, this happens for simply-coupled leads that are attached to the same box.
If two leads cannot be connected via purely bosonic cotunneling processes,
i.e., if a Pauli string is involved,
they must belong to distinct subsectors.
In particular, a lead with a non-simple lead-MZM contact generally defines its own subsector with $M=|{\cal B}|=1$. 
According to this definition, all leads in a general Majorana network uniquely belong to one of its corresponding subsectors.

We start with the case of $M$ leads attached to a given box via 
simple lead-MZM contacts, thus forming a subsector ${\cal B}$. 
In the simplest case, the Hamiltonian describing purely bosonic 
cotunneling processes within this subsector  
follows from Eq.~\eqref{eq:Htunboso} by summing over  all
tunneling paths connecting lead $j\ne k$ (with $j,k\in{\cal B}$).
Such processes have amplitude $J_{j k}$ and couple  different leads only via the 
lead boson fields $\Phi_{j}$ and $\Phi_{k}$.  Adapting Eq.~\eqref{eq:RGgeneral} 
to this purely bosonic problem, we reproduce the RG equations for the single-impurity TKE \cite{Beri2012,Altland2013,Beri2013,Zazunov2014,Altland2014},
\begin{equation}\label{eq:RGTKE}
\frac{dJ_{jk}}{d\ell} = \sum_{ m\in {\cal B}, m \neq (j,k)} J_{j m} J_{mk}.
\end{equation}
 For $M\ge 3$, these couplings automatically scale towards isotropy, $J_{jk}(\ell)\to (1-\delta_{jk}) J(\ell)$,
 see Refs.~\cite{Beri2012,Zazunov2014} for a detailed discussion. The RG equation for the isotropic coupling $J$ is then given
 by $dJ/d\ell=(M-2)J^2$. The isotropic part is thus marginally relevant and flows towards  strong coupling.  
 Deviations from isotropy, on the other hand, are RG irrelevant and can be neglected at low energy scales.
 The TKE thus features an in-built flow to isotropy. The strong-coupling regime is reached at energy scales 
below the Kondo temperature \cite{Beri2012,Altland2013,Beri2013}
\begin{equation}\label{eq:TK}
T_{K} \simeq D e^{-1/[(M-2)\nu J]},
\end{equation}
 where  $D$ is the (bare)
 bandwidth of the leads, and for completeness we re-inserted the lead density of states $\nu$.

Apart from the purely bosonic processes behind Eq.~\eqref{eq:RGTKE},
cotunneling events also can kick the system out of a bosonic subsector  ${\cal B}_1$ into a distinct subsector
${\cal B}_2$, which may belong to the same or to another box.  By definition, such processes 
involve a string $\sigma^1\cdots\sigma^n$ of  $n\ge 1$ Pauli operators. 
The corresponding Hamiltonian reads, cf. Eq.~\eqref{eq:Htunboso2},
\begin{equation}\label{eq:Hbososub2}
H_{\mathrm{nbos}} = \sum_{j\in\mathcal{B}_1}\sum_{k\in\mathcal{B}_2}
J_{j k}^{(\{\sigma\})}\sigma^1\cdots\sigma^n e^{i(\Phi_{j}-\Phi_k)} +\mathrm{h.c.}
\end{equation}
In Appendix \ref{appA}, we illustrate several examples for tunneling processes 
contributing to Eq.~\eqref{eq:Hbososub2} in a rather advanced device with four boxes.
These examples also serve to show the general applicability and versatility of our formalism for arbitrary 
coupled box devices.  

We now study how the RG equations in Eq.~\eqref{eq:RGTKE} 
for purely bosonic couplings $J_{jk}$ with $j\neq k \in {\cal B}$ will be modified by 
the inter-subsector cotunneling processes in Eq.~\eqref{eq:Hbososub2}.  
In general, such an excursion from lead $j\in {\cal B}$ to some other subsector ${\cal B}_2$ must involve a Pauli 
string $\sigma^1\cdots\sigma^n$ with $n\ge 1$.
In order to contribute to the RG flow of our purely bosonic coupling $J_{jk}$, however, the 
tunneling path must now return to lead $k\in {\cal B}$ via the \emph{same} Pauli operator string. 
As a result, for coupled-box networks,  the RG equations for the TKE in Eq.~\eqref{eq:RGTKE} receive an additional
contribution,
\begin{equation}\label{eq:subsectorRG}
\frac{dJ_{jk}}{d\ell} = \sum_{m \in {\cal B},m\ne (j,k)}
J_{j m}J_{mk} +  \sum_{m\notin\mathcal{B}} J_{j m}^{( \{\sigma\})}
 J_{mk}^{(\{\sigma\})}.
\end{equation}
Similarly, see also App.~\ref{appA} for additional details, we obtain the RG equations for the 
cotunneling amplitudes $J_{jk}^{(\{\sigma\})}$, with leads $j\in {\cal B}_1$ and $k\in {\cal B}_2$
belonging to different subsectors, from the general equations \eqref{eq:RGgeneral},
\begin{equation}\label{eq:intersubRG}
\frac{dJ_{jk}^{(\{\sigma\})} }{d\ell} =\sum_{m\in {\cal B}_2, m\ne k} 
J_{jm}^{(\{\sigma\})} J_{mk}+\sum_{m\in {\cal B}_1 ,m\ne j}J_{jm}J_{mk}^{(\{\sigma\})}.
\end{equation}
The first (second) term comprises an inter-sector transition followed by a intra-sector tunneling in $\mathcal{B}_2$ ($\mathcal{B}_1$).
We note that on top of the terms in Eq.~\eqref{eq:intersubRG}, higher-order tunneling
excursions via distinct subsectors $\mathcal{B}'\neq \mathcal{B}_{1,2}$ may generate additional contributions, see App.~\ref{appA}. 
For the applications below, such complications are absent. 

\subsubsection{Anticommuting Pauli strings}

Next we discuss the case of anticommuting Pauli strings $\{\sigma\}$ and $\{\sigma^\prime\}$. 
Using the relation $\mathcal{T}_\tau\sigma_x(\tau)\sigma_y(\tau') = i\sigma_z(\tau)\mathrm{sgn}(\tau-\tau')$
for $\tau\to \tau'$ (and cyclic permutations thereof), with the time-ordering operator ${\cal T}_\tau$, we first 
observe that contributions with different time ordering will interfere destructively.
As a consequence, we find that there will be no additional contributions to the RG equations \eqref{eq:subsectorRG} and \eqref{eq:intersubRG} 
from such tunneling events.

However, other types of RG terms can be generated in systems allowing for closed loops,
where subsectors can be connected through distinct tunneling paths with different Pauli strings.
To that end, let us pick a tunneling path which starts at lead $j\in\mathcal{B}$,
makes an excursion to a lead in some other subsector, $l \notin\mathcal{B}$, and phase-coherently returns 
back to lead $j$. To illustrate the principle,
we here focus on the simplest scenario, where the
Pauli strings $\{\sigma^\prime\}$ and $\{\sigma\}$ for back- and forth-tunneling, respectively, 
are identical except at one link ($m$). At this link, we have anticommuting Pauli operators, say,  
$\sigma_x^m$ and $\sigma_y^m$.
Contracting both cotunneling operators now schematically yields 
\begin{eqnarray}\nonumber
&& J_{jl}^{\left(\ldots\sigma_x^m\ldots\right)}(\cdots\sigma_x^m\cdots)_\tau
  J_{lj}^{\left(\ldots\sigma_y^m\ldots\right)}(\cdots\sigma_y^m\cdots)_{\tau'}
\\  && \sim J_{jl}^{\left(\ldots\sigma_x^m\ldots\right)}
J_{lj}^{\left(\ldots\sigma_y^m\ldots\right)} 
\ i\sigma_z^m(\tau)\mathrm{sgn}(\tau-\tau'),           \label{eq:noncommOPE}
\end{eqnarray}
where all other Pauli operators apart from $\sigma^m_{x,y}$ square out.  
Expanding also the $e^{\pm i\Phi_j}$ factors appearing in all cotunneling operators to lowest order in $\tau-\tau'$, 
we encounter another sgn$(\tau-\tau')$ factor and therefore a finite contribution to the RG equations. 
Using the lead densities near the respective contacts, $\Theta'_j(\tau)=\partial_x\theta_j(x=0,\tau)=-i\partial_{\tau}\phi_j(x=0,\tau)$, see Eq.~\eqref{shorthand}, we then obtain a new contribution generated by such contractions,
\begin{equation}\label{eq:spindensityhyb}
H_\mathrm{hyb} = \sum_{j} \Lambda_{j} \sigma_z^m \Theta_{j}',
\end{equation}
describing a hybridization between $\sigma_z^m$ and the lead fermion densities $\Theta_j'$.  
(Of course, depending on the application, the coupling in Eq.~\eqref{eq:spindensityhyb} may involve other or even multiple Pauli operators.)
We note that similar terms also appear in the context of charge Kondo effects \cite{Emery1992,Gogolin1998,Landau2017b}.

From Eq.~\eqref{eq:noncommOPE}, the RG flow of the coupling constants in Eq.~\eqref{eq:spindensityhyb} is then governed by
\begin{equation}\label{eq:spindensityhybRGa}
\frac { d\Lambda_{j}}{d\ell} 
\sim \sum_{l\notin\mathcal{B}} J_{jl}^{\left(\ldots\sigma_x^m\ldots\right)}
J_{lj}^{\left(\ldots\sigma_y^m\ldots\right)}+\mathrm{h.c.}
\end{equation}
Hybridization couplings thus will be dynamically created during the RG flow even for vanishing bare coupling,
i.e., for $\Lambda_j(\ell= 0)=0$.  We remark in passing that 
$\Lambda_j(0)\neq 0$ could arise from in- and out-tunneling events at a lead contacting several MZMs, cf.~Fig.~\ref{fig2}(b).
The $\Lambda_{j}$ are real-valued couplings which are effectively controlled by the sine or cosine of the loop phase
\begin{equation}\label{eq:loophase}
\varphi^\mathrm{loop}_{j} = \arg\left(\sum_{l\notin\mathcal{B}}
J_{jl}^{\left(\ldots\sigma_x^m\ldots\right)}J_{lj}^{\left(\ldots\sigma_y^m\ldots\right)}\right).
\end{equation}
Importantly, the hybridizations in turn feed back into the RG equations \eqref{eq:intersubRG} for cotunneling amplitudes.
In fact, we find that Eq.~\eqref{eq:intersubRG} receives the additional contributions 
\begin{equation}\label{eq:feedbackRGx}
\frac{dJ_{j l}^{\left(\ldots\sigma_{x/y}^m\ldots\right)}}{d\ell}\sim\left(\Lambda_{l}-\Lambda_{j}\right)
J_{j l}^{\left(\ldots\sigma_{y/x}^m\ldots\right)}.
\end{equation}
For the loop qubit example studied below, see Secs.~\ref{sec3d} and \ref{sec4e}, such RG feedback effects turn out to be crucial.

\subsubsection{Summary}

The above rules show that RG equations for a general 
coupled Majorana box system can be determined by contracting pairs of tunneling operators.  Commuting tunneling operators  generate 
new composite tunneling operators and/or renormalize existing couplings, see
Eqs.~\eqref{eq:subsectorRG} and \eqref{eq:intersubRG}. 
Contractions of non-commuting operators, on the other hand, do not contribute to the latter RG equations. However, in systems with tunneling paths forming closed loops,  hybridization terms between Pauli operators and lead fermion densities will be generated.  Such terms will in turn feed back into the RG equations for the cotunneling amplitudes.
Next we apply the above RG analysis to several examples of practical interest.

\subsection{Two-box device}\label{sec3b}

Let us begin by studying a two-box device as shown in Fig.~\ref{fig3}. 
We first observe that such a system does not admit tunneling paths forming closed loops, 
and thus the RG equations do not involve the hybridizations in Eq.~\eqref{eq:spindensityhyb}. 
Using $H_{\rm leads}$ in Eq.~\eqref{eq:Hleads} and
taking into account the central junction described by
Eq.~\eqref{eq:Htuntwobox}, the Hamiltonian $H=H_{\rm leads}+H_L+H_R+H_{LR}$ 
is obtained by a Schrieffer-Wolff transformation to the ground-state charge sector of both boxes, see Sec.~\ref{sec2d}.  
 In particular, cotunneling processes involving only boson fields connected to the left/right ($L/R$) box are contained in
 \begin{eqnarray}\label{eq:HtwoboxL}
 H_{L/R}&=& -\sum_{j,k\in\mathcal{B}_{L/R},j\ne k} (J_{L/R})_{jk} \cos\left(\Phi_j-\Phi_k\right)
\\ \nonumber &-& \sum_{j\in\mathcal{B}_{L/R}} (J_{X})_{l/r,j} 
\sigma_x\cos\left(\Phi_{l/r}-\Phi_j\right), 
\end{eqnarray}
where ${\cal B}_{L/R}$ denotes bosonic subsectors with $M_{L/R}$ leads connected to the respective box via simple lead-MZM contacts.
(For the example in Fig.~\ref{fig3}, $M_L=3$ and $M_R=2$.)
The central leads in Fig.~\ref{fig3}, with boson fields $\Phi_{l/r}$, are coupled to 
the $L/R$ box via non-simple contacts, where  
non-conserved local fermion parities are encoded by the Pauli operators $\sigma_{x,y,z}$, see Eq.~\eqref{eq:Htuntwobox}.
Inter-box cotunneling processes are contained in 
\begin{eqnarray}\label{eq:HtwoboxLR}
H_{LR}&=&-\sum_{j\in{\cal B}_L} (J_{Y})_{rj}\ \sigma_y\cos\left(\Phi_r-\Phi_{j}\right) \\ \nonumber
&-& \sum_{k\in{\cal B}_R} (J_{Y})_{l k} \ \sigma_y\cos\left(\Phi_l-\Phi_{k}\right) \\ \nonumber
&+&\sum_{j\in{\cal B}_L,k\in{\cal B}_R } (J_{Z})_{jk}\ \sigma_z \sin\left(\Phi_{j}-\Phi_{k}\right).
\end{eqnarray}
The $J_{L/R}$ amplitudes in Eq.~\eqref{eq:HtwoboxL} are purely bosonic intra-sector couplings as in Sec.~\ref{sec3a}.
The $J_X$  (resp., $J_Y$) cotunneling amplitudes connect leads within bosonic subsector ${\cal B}_{L/R}$ 
to the central lead on the same (resp., other) box, involving the Pauli string $\sigma_{x}$ (resp., $\sigma_y$).  
Finally, the $J_Z$ amplitudes link the bosonic subsectors ${\cal B}_L$ and ${\cal B}_R$
by inter-box tunneling via the Pauli string $\sigma_z$.

In total, we thus have seven coupling families: $J_{L/R}, J_{X,l/r}, J_{Y,r/l}$, and $J_Z$.
The respective coupling matrix elements depend on microscopic lead-MZM ($\lambda_j$) and  
MZM-MZM ($t_{LR}$) tunneling amplitudes, cf.~Eq.~\eqref{eq:Htuntwobox}. Schematically, 
 $(J_{L/R/X})_{jk}\sim \lambda_j \lambda_k^\ast / E_C$ and 
$(J_{Y/Z})_{jk}\sim \lambda_j \lambda_k^\ast t_{LR}/E_C^2$.
Since one can gauge away complex phases of tunneling amplitudes for systems without closed loops,  
all these cotunneling amplitudes can be chosen real positive. Within each coupling family, we thus arrive at a real 
symmetric matrix. 

The RG equations then follow from 
Eqs.~\eqref{eq:subsectorRG} and \eqref{eq:intersubRG}.  For $j,k\in {\cal B}_L$, we  find
\begin{eqnarray}\nonumber
\frac{d(J_{L})_{j k}}{d\ell} &=& \sum_{m\in{\cal B}_L, m\neq (j,k)}
(J_L)_{j m} (J_L)_{mk} +(J_X)_{lj} (J_X)_{lk} \\ \label{eq:RGtwobox1}
&+&(J_Y)_{rj} (J_Y)_{rk}+\sum_{m\in{\cal B}_R} (J_Z)_{jm}(J_Z)_{mk} .
\end{eqnarray}
Furthermore, with $j\in {\cal B}_L$, we get
\begin{equation}
\frac{d(J_{X/Y})_{l/r,j}}{d\ell}=
\sum_{m\in{\cal B}_L,m\neq j }(J_{X/Y})_{l/r,m}(J_L)_{m j},
\end{equation}
while for $j\in{\cal B}_L$ and $k\in {\cal B}_R$,  
\begin{eqnarray} \nonumber
\frac{d(J_Z)_{jk}}{d\ell}&=&\sum_{m\in{\cal B}_L,m\neq j} (J_L)_{j m}(J_Z)_{mk}
\\ &+&\sum_{m\in{\cal B}_R,m\neq k} (J_Z)_{jm}(J_R)_{mk}.
\label{eq:RGtwobox4}
\end{eqnarray}
The corresponding RG equations for the $J_R$, $J_{X,r}$ and $J_{Y,l}$ couplings  follow by exchanging left/right labels.

The above RG equations can be simplified considerably by observing that different coupling families 
 effectively become isotropic at low energy scales.  
For small-to-moderate bare anisotropies of the respective coupling matrices, such an isotropization
can already be established within the weak-coupling regime accessible to the RG approach.
For the single-box TKE case with $M \geq 3$ leads, this mechanism has been detailed in Refs.~\cite{Beri2012,Altland2013,Beri2013,Zazunov2014}.
As shown in App.~\ref{appB} by a numerical solution of the full RG equations
\eqref{eq:RGtwobox1}--\eqref{eq:RGtwobox4}, the isotropization mechanism also applies for the 
two-box device in Fig.~\ref{fig3} with $M_R=2$. By a similar analysis, we have verified that isotropization 
applies for all other examples where we invoke it below.
This finding can be rationalized by noting that for any $M\ge 2$, couplings to leads in this sector feed back into the RG flow of each other if they belong to the same family.
As a consequence, different coupling families are effectively described by specifying only
their mean (average) values, $(J_L)_{jk}\to J_L$ and so on, see Eq.~\eqref{eq:appB1} in App.~\ref{appB}.
Anisotropies within a given coupling family are  RG irrelevant and thus can be neglected at low energies.
In fact, we expect the above conclusions to apply for general coupled Majorana box systems.

The two-box problem in Fig.~\ref{fig3} is then described by seven running couplings, where 
Eqs.~\eqref{eq:RGtwobox1}--\eqref{eq:RGtwobox4} yield the isotropized RG equations
\begin{eqnarray}
\frac{dJ_L}{d\ell} &=&(M_L-2)J_L^2 +M_RJ_{Z}^2+ J_{X,l}^2+J_{Y,r}^2,\nonumber\\
\frac{dJ_{X,l}}{d\ell}&=&(M_L-1)J_{X,l} J_L, \quad
\frac{dJ_{Y,r}}{d\ell}=(M_R-1)J_{Y,r}J_L, \nonumber\\
\frac{dJ_Z}{d\ell}&=& \left[(M_L-1)J_L+(M_R-1)J_R\right] J_Z,
\label{eq:RGtwoboxiso}
\end{eqnarray}
and related equations for $J_{R}$, $J_{X,r}$, and $J_{Y,l}$.
Let us briefly check Eq.~\eqref{eq:RGtwoboxiso} for two limiting cases: 
\begin{enumerate}[(i)]
\item
For vanishing MZM-MZM coupling, $t_{LR}\to 0$, both boxes are decoupled. We thus have 
$J_{Z}=J_{Y,r/l}= 0$, and $\sigma_x=\pm 1$ is conserved. The above equations then reduce to a decoupled pair of 
single-impurity TKE systems, cf.~Eq.~\eqref{eq:RGTKE}, where $M_L+1$ and $M_R+1$ leads are attached
 to the left/right box: for $t_{LR}=0$, the central leads $l$ and $r$ in Fig.~\ref{fig3} join the 
 respective bosonic subsector ${\cal B}_{L/R}$.
 \item In the absence of both central leads, we have $J_{X,l/r}=J_{Y,r/l}= 0$ and
$\sigma_z = \pm$ is conserved. In that case, we recover the RG equations for
 the single-impurity TKE again. However, since both boxes are now connected by $t_{LR}\ne 0$, 
 we encounter the equations for a \emph{single} Kondo impurity with $M_L+M_R$ attached leads. At low energies, both boxes
 are thus fused together by the MZM-MZM link and thereby form a single enlarged Majorana box that subsequently
 exhibits a global TKE with symmetry group SO$_2(M_L+M_R)$.
\end{enumerate}
For generic initial values of the isotropized cotunneling amplitudes, we have numerically solved the 
RG equations \eqref{eq:RGtwoboxiso}.  Our analysis shows that the system will flow towards strong coupling with
competing separate (intra-box) and global (inter-box) TKEs. This scenario is reminiscent of the classic two-impurity Kondo problem \cite{Jayprakash1981,Jones1988,Affleck1995}
and indicates that a strong-coupling analysis is needed in order to determine the ground state, see Sec.~\ref{sec4c}.

\subsection{MZM coupled to multiple leads}
\label{sec3c}

An interesting limit of the two-box RG equations \eqref{eq:RGtwoboxiso} concerns the physics of a single MZM 
coupled to several leads, see Fig.~\ref{fig2}(c) and Eq.~\eqref{eq:Htun12}. To this end, one may consider
a situation where the left (resp., right) box has $M\equiv M_L$  (resp., $M_R=1$) leads with simple lead-MZM contacts.  These leads are described by the boson fields $\Phi_{j\in{\cal B}_L}$ (resp., $\Phi_z$).
We then note that the MZM $\gamma_l$ on the left box, which is tunnel-coupled to the central lead $\Phi_x\equiv\Phi_l$ in Fig.~\ref{fig3}, effectively also couples to the two leads connected to the right box via the MZM-MZM tunnel bridge. 
Let us write $\Phi_y\equiv\Phi_r$ for the corresponding central lead and use isotropic couplings for 
different coupling families, see Sec.~\ref{sec3b}, where isotropization holds for $M\ge 2$.  Retaining for
the moment only the four couplings
\begin{equation}\label{fourcot}
J=J_L, \quad J_x=J_{X,l},\quad J_y=J_{Y,r},\quad J_z=J_Z,
\end{equation}
 the low-energy Hamiltonian is $H=H_{\rm leads}+H_b$, with the boundary term
\begin{eqnarray}\label{eq:HtunB}
H_b &=& -J\sum_{j,k \in {\cal B}_L,j\ne k} \cos(\Phi_j-\Phi_k) \\ \nonumber
&-& \sum_{\alpha=x,y,z} J_\alpha\sigma_\alpha \sum_{j\in{\cal B}_L} \cos(\Phi_j-\Phi_\alpha).
\end{eqnarray}
The $J_\alpha$ in Eq.~\eqref{fourcot} thus characterize our lead-MZM multi-junction.
We emphasize that the right box in the above setup is not necessary for observing the physics below,
and one could simply couple the leads corresponding to the fields $\Phi_{x,y,z}$ directly to $\gamma_l$.
Its inclusion here only allows us to take over results from Sec.~\ref{sec3b}.

In fact, the corresponding RG equations can now be read off from Eq.~\eqref{eq:RGtwoboxiso},
\begin{equation}
\frac{dJ}{d\ell} =(M-2)J^2 +\sum_{\alpha} J_\alpha^2 ,\quad \frac{dJ_\alpha}{d\ell}=(M-1)JJ_\alpha.\label{eq:RGmultlead}
\end{equation}
Cotunneling processes between  the three leads $\Phi_{x,y,z}$ are 
not contained in Eq.~\eqref{eq:HtunB} and arise due to the three remaining 
couplings $J_R$, $J_{X,r}$ and $J_{Y,l}$ beyond those in Eq.~\eqref{fourcot}.  
Such terms generate the additional contribution 
$H'_b\sim\sigma_z\cos(\Phi_x-\Phi_y)$ plus cyclic permutations.  From the analysis in Sec.~\ref{sec2d}, we find $H'_b=0$ under 
Coulomb valley center conditions, i.e., for $\Delta n_g=0$.   In any case, 
such couplings are neither RG relevant, in contrast to those in Eq.~\eqref{fourcot}, nor do 
they enter the flow of other couplings in Eq.~\eqref{eq:RGmultlead}. We can thus safely drop them in what follows.

Let us then discuss the RG flow generated by Eq.~\eqref{eq:RGmultlead}.
First, we observe that  ratios of different $J_\alpha$ couplings  are conserved,
$dJ_x/dJ_y = J_x(0)/J_y(0)$ and $dJ_y/dJ_z = J_y(0)/J_z(0)$. 
All $J_\alpha$ therefore flow towards strong coupling together with those ratios being invariant.
Second, for $M\geq 3$, the TKE-like coupling $J$ outgrows the $J_\alpha$ 
since they all feed back into the RG flow of $J$.   In contrast, for $M=2$, we observe that $J$ does not benefit
 from the self-enhanced TKE-like RG flow, cf.~Eq.~\eqref{eq:RGTKE}, and therefore will not 
 automatically dominate anymore.  In fact, for $M=2$, Eq.~\eqref{eq:RGmultlead} becomes a 
multi-component version of the celebrated Kosterlitz-Thouless equations \cite{Altland2010}, where
the RG flow of $J$ is directly induced by the $J_\alpha$ flow and vice versa.
In our strong-coupling analysis of this setup, see Sec.~\ref{sec4d}, 
we will focus on the most interesting case $M=2$. 
Further transport properties for this system are discussed in Sec.~\ref{sec5}.

\subsection{Loop qubit}\label{sec3d}

\begin{figure}
\centering  
\includegraphics[width=0.45\textwidth]{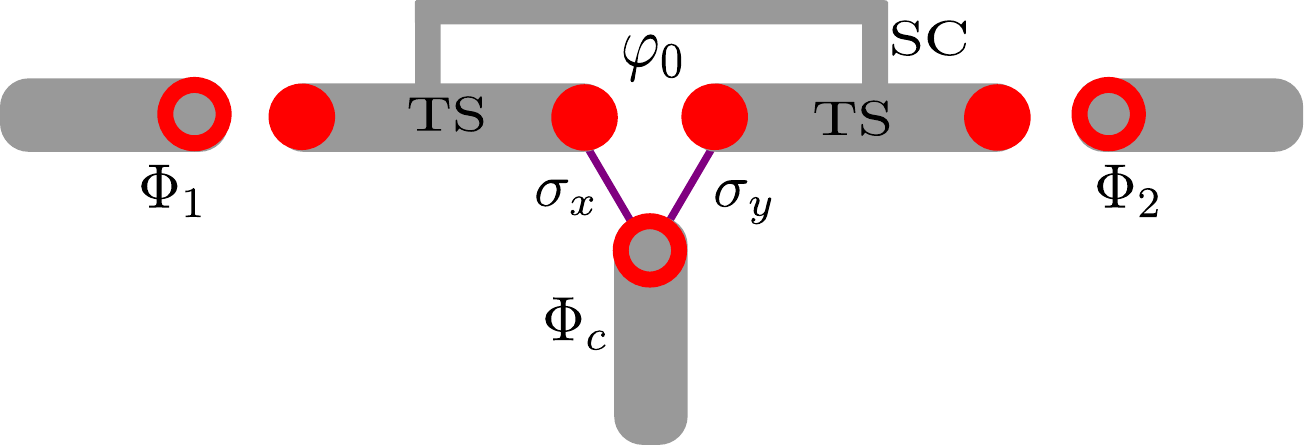}
\caption{
Loop qubit device contacted by normal leads. This device has been suggested  in Fig.~14 of Ref.~\cite{Karzig2017} 
for interferometric Majorana qubit measurements and manipulations, see also Refs.~\cite{Landau2016,Plugge2016c,Plugge2017}.  
Two long topological superconductor (TS) wires with a superconducting (SC) bridge define a Majorana box with four MZMs, 
where the loop phase $\varphi_0$ can be controlled by a magnetic flux.  The normal leads attached to the box
correspond to boson fields $\Phi_{1,2,c}$. The central lead ($\Phi_c$) couples to two MZMs as in Fig.~\ref{fig2}(b),
where the non-conserved fermion parity is encoded by Pauli operators $\sigma_{x,y,z}$.
 For an explanation of symbols, see Figs.~\ref{fig1} and \ref{fig2}. }
	\label{fig4}
\end{figure}

As final example for the RG analysis, we here consider the loop qubit device shown in Fig.~\ref{fig4}.
This device has a single Majorana box containing $M=2$ leads with simple contacts, and
a non-simple contact coupling two MZMs to a central lead (with boson field $\Phi_c$), see Sec.~\ref{sec2c}, in particular 
Eq.~\eqref{eq:Htun21} and Fig.~\ref{fig2}(b). 
Importantly, such a device provides the simplest possibility for tunneling paths forming closed loops.  
It has been suggested as Majorana qubit realization \cite{Karzig2017}, 
where the relative phase $\varphi_0$ between the tunneling amplitudes connecting the central lead
with the respective MZM can be changed by a magnetic flux. We note
that $\varphi_0$ corresponds to the loop phase between different tunneling paths in Eq.~\eqref{eq:loophase}.
By contacting the box with leads as shown in Fig.~\ref{fig4}, nontrivial  interferometric conductance 
measurements can be performed.  In particular, a measurement of the linear conductance between 
the central lead and one of the outer leads ($\Phi_{1,2}$ in Fig.~\ref{fig4}) could
determine the eigenvalue of the Pauli operator $\sigma_z$ related to 
the non-conserved fermion parity of the junction \cite{Plugge2017,Karzig2017}.

The non-simple junction is described by $H_{2,1}$ in Eq.~\eqref{eq:Htun21} with $\Phi\to \Phi_c$ and $h_z\to 0$.
We do not include a direct MZM-MZM coupling, but MZMs instead hybridize with the fermion density at the central contact, see below.
With $\sigma_\pm=(\sigma_x\pm i\sigma_y)/2$, we thus have 
\begin{eqnarray}\nonumber
H_{2,1} &=& (\lambda_+\sigma_++\lambda_-\sigma_-)e^{i(\varphi-\Phi_c)}+\mathrm{h.c.},\\
\lambda_\pm &=& \lambda_x\mp i\lambda_y e^{i\varphi_0}, \label{eq:HtunC}
\end{eqnarray}
where we use a gauge where $\varphi_0$ appears
at the $\sigma_y$ link in Fig.~\ref{fig4} and the tunneling amplitudes $\lambda_{x,y}$ are real-valued. 
Interestingly, for $\varphi_0=\pi/2$, the same model describes quasi-particle poisoning effects for the TKE 
 \cite{Plugge2016b}. 

As next step, we implement the projection to the ground-state charge of the box, see  Sec.~\ref{sec2d}.
Following the corresponding steps in Ref.~\cite{Plugge2016b} but allowing for arbitrary loop phase $\varphi_0$,
we then get the Hamiltonian $H=H_{\rm leads}+H_b$. For $M$ leads (labeled by $j\in {\cal B}$) with simple 
 contacts to the box, where $M=2$ in Fig.~\ref{fig4}, 
\begin{eqnarray}\nonumber
H_b &=& -J \sum_{j,k \in {\cal B}, j\ne k}\cos\left(\Phi_j - \Phi_k\right) -\sum_{j\in {\cal B}}
\tilde{\Lambda}\sigma_z\Theta'_j - \Lambda_c\sigma_z\Theta'_c \\
&-&\frac{1}{\sqrt2}\sum_{j\in{\cal B}} \left[(L_+\sigma_+ + L_-\sigma_-)e^{i(\Phi_j - \Phi_c)} +
{\rm h.c.}\right], \label{eq:Hqpp}
\end{eqnarray}
where we assume isotropic couplings. With a tunnel coupling $\tilde \lambda$ for the simple lead-MZM contacts,
the complex-valued cotunneling amplitudes between the central and the outer leads are contained in 
$L_\pm =\sqrt2 \tilde\lambda\lambda_\pm/E_C$, see Eq.~\eqref{eq:HtunC}.  In contrast, the TKE-like coupling $J$ describes cotunneling
between leads within subsector ${\cal B}$.  Because of the existence of tunneling paths
forming closed loops, Eq.~\eqref{eq:Hqpp} also contains hybridization terms of the form 
in Eq.~\eqref{eq:spindensityhyb}. The bare (initial) values for these couplings are  $\tilde{\Lambda} = 0$ and 
$\Lambda_c \simeq (\lambda_x\lambda_y/E_C)\sin\varphi_0$. 
During the RG flow, both $\tilde{\Lambda}$ and $\Lambda_c$ grow and approach strong coupling.

We next exploit current conservation, 
$\braket{\Theta_c'}+\sum_j\braket{\Theta_j'} = 0$, which follows from gauge invariance under a 
simultaneous shift of all boson fields $\Phi_{j,c}$.
This relation allows us to further reduce the number of parameters by trading off hybridizations at the outer leads versus an 
enhanced hybridization between the central lead and $\sigma_z$. 
With $\Lambda = 2(\Lambda_c-\tilde{\Lambda})$, we then obtain the RG equations, cf.~Ref.~\cite{Plugge2016b},
\begin{eqnarray}\nonumber
\frac{dJ}{d\ell} &=& (M-2)J^2+\vert L_+\vert^2+\vert L_-\vert^2,\\
\label{eq:RGqpp} 
\frac{dL_{\pm}}{d\ell} &=&\left[(M-1)J\pm \Lambda\right]L_\pm,\\
\nonumber  \frac{d\Lambda}{d\ell} &=&(M+1)\left(\vert L_+\vert^2-\vert L_-\vert^2\right).
\end{eqnarray}
The most interesting prediction of these equations is the onset of \emph{helicity} \cite{Plugge2016b}, i.e., a nontrivial 
flow of the couplings $L_\pm$. 
To this end, it is instructive to relate the RG flow of the above couplings with that of the loop phase $\varphi_0$.
We first observe that with $\lambda_\pm$ in Eq.~\eqref{eq:HtunC}, 
\begin{eqnarray}
\vert L_+(\ell)\vert^2+\vert L_-(\ell)\vert^2 &\sim& \lambda_x^2+\lambda_y^2, \nonumber\\
\vert L_+(\ell)\vert^2-\vert L_-(\ell)\vert^2 &\sim& \lambda_x\lambda_y\sin\varphi_0.
\end{eqnarray}
This implies that while the TKE-like coupling $J$ grows and stays independent
 of $\varphi_0$, the hybridization $\Lambda$, with initial value $\Lambda(\ell = 0) \sim 
 \sin\varphi_0$, keeps the same dependence on $\varphi_0$ throughout the RG flow. 
Moreover, the complex phases of the couplings $L_\pm$ are invariant during the RG flow 
since the prefactor for their self-renormalization in Eq.~\eqref{eq:RGqpp} is real. 
Using $L_\pm\sim\lambda_\pm$, the running loop phase is then defined by 
\begin{equation}\label{eq:RGloopphase}
\varphi_0(\ell) = \arg\left[i(L_+-L_-)/(L_++L_-)\right]_\ell.
\end{equation}
Note that $\varphi_0(\ell)$ will in general change during the RG flow because it depends on both the 
complex phases and the absolute values of $L_\pm$.
In particular, we observe that for bare loop phases with $\varphi_0(0)\in (0,\pi)$, we 
will also have $|L_+(0)|>|L_-(0)|$, while for $\varphi_0(0)\in (-\pi,0)$, we instead find $|L_-(0)|>|L_+(0)|$. 
The  RG equations \eqref{eq:RGqpp} thus predict a flow of the bigger coupling $L_\pm$ to strong coupling, along with 
growing $J$ and $\Lambda$, while the opposite coupling $L_\mp$ is dynamically suppressed.

\begin{figure}
\centering  
\includegraphics[width=0.4\textwidth]{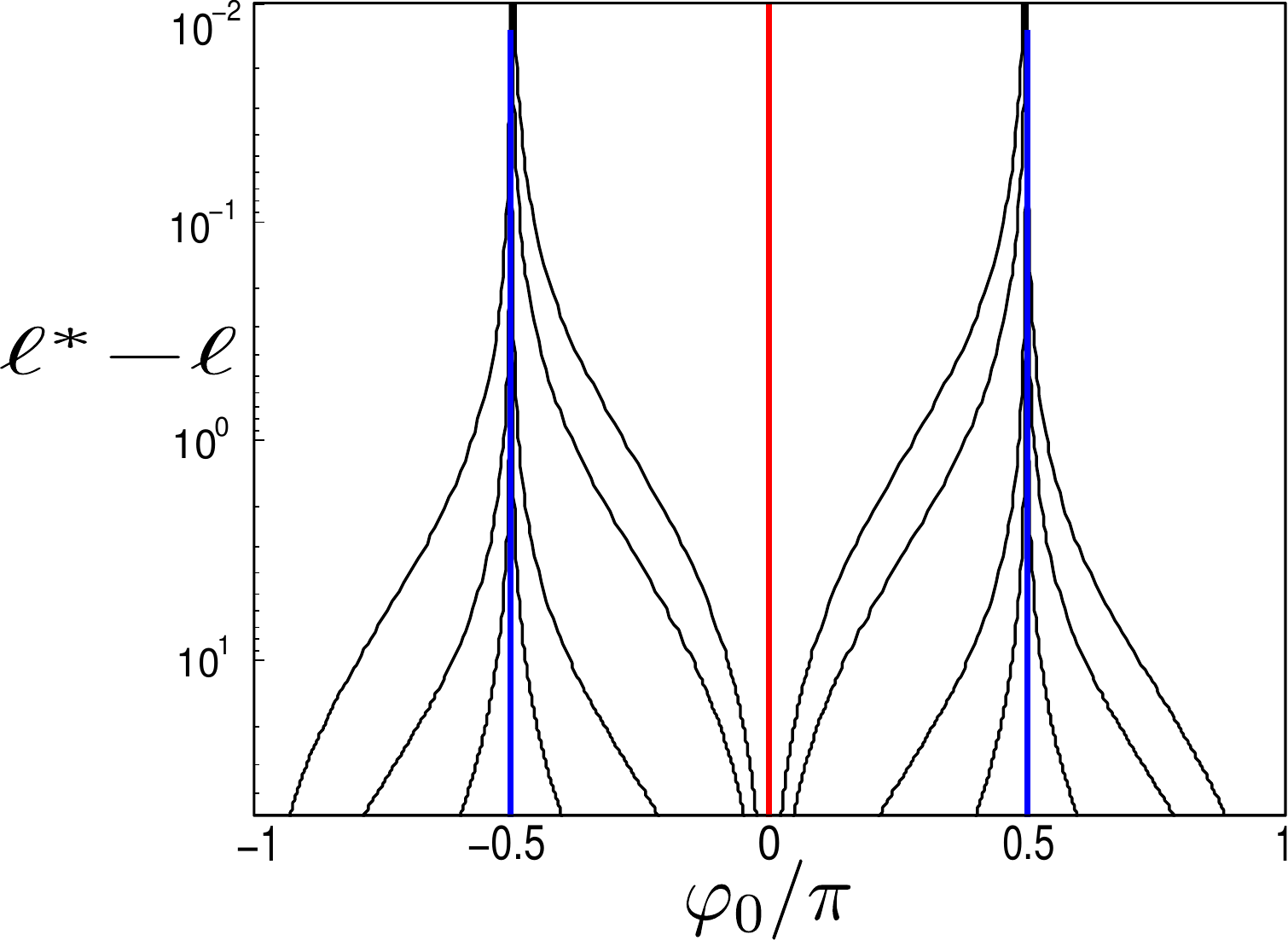}
\caption{RG flow of the loop phase $\varphi_0(\ell)$ obtained by numerical integration of a fully anisotropic version \cite{Plugge2016b} of the RG equations \eqref{eq:RGqpp}.  At $\ell=\ell^*$, the RG approach breaks down due to divergent
couplings. We show results for  $\ell^*-\ell$ vs $\varphi_0$ on a semi-logarithmic scale.
  While this plot was generated for a specific randomly chosen
set of initial parameters, with different $\varphi_0(0)$, we have checked that the qualitative features of the RG flow are insensitive to this choice.  
We identify two stable fixed points with $\varphi_0=\varphi_\pm = \pm\pi/2$ (blue vertical lines),
where the hybridization is maximized, $\Lambda(\ell)\sim \sin\varphi_0(\ell)$. In contrast, $\varphi_0 = 0$~mod~$\pi$ (red vertical line) are unstable fixed points with $\Lambda=0$.
The gauge symmetry of the system under the exchange $L_+\leftrightarrow L_-$, $\sigma_+\leftrightarrow\sigma_-$ and $\sigma_z\to-\sigma_z$, 
cf.~Eqs.~\eqref{eq:Hqpp} and \eqref{eq:RGqpp}, is apparent in the symmetry of the
RG flow under $\varphi_0\to - \varphi_0$.}
\label{fig5}
\end{figure}

In Fig.~\ref{fig5}, we show typical results for the RG flow of $\varphi_0$ obtained by numerical integration of 
a fully anisotropic version of Eq.~\eqref{eq:RGqpp}.
The numerical results perfectly recover the qualitative behavior discussed above. We note that these calculations have
also confirmed that all couplings indeed become isotropic during the RG flow.
In physical terms, the limiting cases of the RG flow in Fig.~\ref{fig5} correspond to phase pinning at low energies, with the 
stable asymptotic value $\varphi_\pm = \pm\pi/2$ as $L_\pm$ outgrows $L_\mp$, cf.~Eq.~\eqref{eq:RGloopphase}.
These two values correspond to the helical fixed points found  in Ref.~\cite{Plugge2016b}. 

Instead, for $\varphi_0 = 0$ or $\varphi_0=\pi$, the RG flow of the hybridization, $\Lambda(\ell)\sim\sin\varphi_0(\ell)=0$, is
fully blocked. Remarkably, in terms of $J_x = (L_++L_-)/2$ and $J_y = -i(L_+-L_-)/2$, we now recover the RG equations 
\eqref{eq:RGmultlead} for the fundamentally different problem of a single MZM coupled to two leads.
These flow equations (with $J_z=0$) imply a flow to strong coupling of $J$ and $J_{x,y}$, 
with fixed ratio $J_x/J_y$, see Sec.~\ref{sec3c}.  

We will return to the loop qubit device in our discussion of the strong-coupling limit in Sec.~\ref{sec4e}.

\section{Strong-Coupling Regime}
\label{sec4}

In Sec.~\ref{sec3} we have seen that, in general, the systems studied here will approach the strong-coupling
regime.  At very low energies, in particular for an understanding of the ground state, one therefore has to go beyond the RG approach.
In this section, we extend concepts developed for a strong-coupling solution of the TKE via Abelian bosonization
 \cite{Beri2013,Altland2013,Zazunov2014,Altland2014,Eriksson2014,Beri2017} to our more general setting.
Such strategies can lead to additional insights, and  even allow for analytical solutions in not too complicated setups.
 
The arguments in Sec.~\ref{sec3} imply that at low energy scales, we need to keep only isotropic cotunneling amplitudes
within and in between subsectors.  In fact, if a subsector contains more than one lead, 
the center-of-mass field will be the only linear combination that is 
not pinned in the ground state. To access the ground state, we 
thus need to study the combined dynamics of these center-of-mass fields and the Pauli operator strings in 
the system.  In this way, 
the complexity of the problem can be drastically reduced and the physics becomes more transparent, see Sec.~\ref{sec4a}. 
A second key ingredient of our strong-coupling approach is tied to the possibility of decoupling certain linear 
combinations of boson fields via unitary transformations, see Sec.~\ref{sec4b}.  We illustrate this strategy 
in Secs.~\ref{sec4c}--\ref{sec4e} for the three applications discussed from the RG viewpoint in Secs.~\ref{sec3b}--\ref{sec3d}.

\subsection{Reduction of bosonic subsectors}\label{sec4a}

Our first step in the construction of the strong-coupling theory is the reduction of every bosonic subsector ${\cal B}$
 to the corresponding center-of-mass field,  
\begin{equation}\label{eq:comphase}
\phi_0(x,\tau) = g_0\sum_{j\in{\cal B}}\phi_{j}(x,\tau), \quad g_0= \frac{1}{\sqrt{M}},
\end{equation}
where $\Phi_0=\phi_0(x=0)$.
For $M=1$, the field $\Phi_0$ then just coincides with the single boson field in the respective subsector (with $g_0=1$),
but Eq.~\eqref{eq:comphase} implies a reduction of complexity for $M=|{\cal B}|\ge 2$.
The  usefulness of Eq.~\eqref{eq:comphase} follows from previous Abelian bosonization studies of 
the strong-coupling TKE \cite{Altland2013,Beri2013,Beri2017,Herviou2016,Michaeli2017} and from our arguments
in Sec.~\ref{sec3}.  In fact, for $M\ge 2$, couplings within ${\cal B}$ grow strong, and for $M\geq 3$ also become
 isotropic.  (However, isotropy is not necessary for our discussion below.)
In detail, following Refs.~\cite{Altland2013,Beri2013,Zazunov2014}, we introduce
 reduced boson fields, $\tilde{\Phi}_{j\in {\cal B}} = \Phi_{j}-g_0\Phi_0$, with the
constraint $\sum_j\tilde{\Phi}_{j} = 0$.
Next, we recall that intra-subsector cotunneling amplitudes $J_{jk}$ (with $j,k\in{\cal B}$) can
be chosen real positive upon absorbing tunnel phases into lead phase fields.
We hence obtain the Hamiltonian for the subsector as 
\begin{equation}\label{eq:reduce}
H_{\mathcal{B}} = - \sum_{j,k\in\mathcal{B}, j\ne k}J_{jk}\cos(\tilde{\Phi}_{j}-\tilde{\Phi}_{k}).
\end{equation}
At strong coupling, the low-energy physics in $\mathcal{B}$ exhibits 
an analogy to the quantum Brownian motion  of a particle with coordinates $\tilde{\Phi}_{j}$ in the $(M-1)$-dimensional lattice defined by the potential $H_{\mathcal{B}}$ \cite{Yi1998,Yi2002,Altland2013,Beri2013,Zazunov2014}.
The motion along the $\Phi_0$-direction is analogous to that of a free particle with linear dispersion,
inherited from the free boson theory in Sec.~\ref{sec2b}. In particular,  
Eq.~\eqref{eq:reduce} does not introduce an energy cost along this direction.
The free field $\Phi_0$ thus dominates the low-energy physics. The leading irrelevant operators at the strong-coupling 
fixed point then come from tunneling events connecting neighboring lattice minima \cite{Yi1998,Yi2002},
 corresponding to quantum phase slips between static configurations $\lbrace\tilde{\Phi}_{j}\rbrace \equiv \lbrace\tilde{\varphi}_{j}\rbrace$ minimizing  $H_{{\cal B}}$  
under the constraint  $\sum_j \tilde\varphi_j=0$.
Such phase slips can be triggered  by electron-hole pair excitations (causing Ohmic dissipation) in the leads \cite{Altland2013,Beri2013},  or due to an applied bias voltage \cite{Beri2017}. 
In fact, scaling dimensions of non-Fermi liquid corrections at the strong-coupling point can be obtained by a geometric analysis of the lattice potential in Eq.~\eqref{eq:reduce} \cite{Yi1998,Yi2002}.
 
 Our main interest in this work is not in effects caused by such intra-subsector leading irrelevant operators. Instead, we 
want to clarify how different center-of-mass boson fields in a coupled box device  interact among themselves 
and with Pauli string operators.
We thus assume that all reduced fields in bosonic subsectors are pinned to their static quasi-classical minima, and
then express the dynamics of $\Phi_{j}$ in terms of the center-of-mass motion,  
\begin{equation}\label{eq:phasefreeze}
\Phi_{j\in {\cal B}}(\tau) = \tilde{\varphi}_{j}+g_0\Phi_0(\tau).
\end{equation}
We note that Eq.~\eqref{eq:phasefreeze} is  appropriate for ground-state properties but misses the leading irrelevant operators discussed above.
However, their effects are quite well understood
and in any case could be added \textit{a posteriori} via perturbation theory.
Let us now consider the effects of the projection in Eq.~\eqref{eq:phasefreeze} on inter-subsector coupling terms.
Inserting Eq.~\eqref{eq:phasefreeze} into Eq.~\eqref{eq:Hbososub2}, 
for transitions between subsectors ${\cal B}_1$ and ${\cal B}_2$, we find the term
\begin{equation}\label{eq:ab}
H_{\mathcal{B}_1,\mathcal{B}_2} = \sum_{j\in\mathcal{B}_1}\sum_{k\in\mathcal{B}_2}
J_{jk}^{(\{\sigma\})}\sigma^1\cdots\sigma^n  e^{i(\tilde{\varphi}_{j}-\tilde{\varphi}_{k})}
e^{i(g_1\Phi_1-g_2\Phi_2)},
\end{equation}
where $\Phi_{1,2}$ denote the center-of-mass fields for subsectors ${\cal B}_{1,2}$, respectively, with $g_{1,2}$ in Eq.~\eqref{eq:comphase}.

Since in Eq.~\eqref{eq:reduce} we gauged away relative tunnel phases between leads in each subsector, 
the $J_{jk}^{(\{\sigma\})}$ in Eq.~\eqref{eq:ab} are real positive up to a global inter-sector phase $\varphi_{{\cal B}_1 {\cal B}_2}^{(\{\sigma\})}$.
Defining an effective tunneling amplitude between sectors $\mathcal{B}_1$ and $\mathcal{B}_2$ 
with the corresponding Pauli string $\{\sigma\}$, 
\begin{equation}\label{eq:interseccoupling}
J_{{\cal B}_1{\cal B}_2}^{(\{\sigma\})} = e^{i\varphi_{{\cal B}_1{\cal B}_2}^{(\{\sigma \})}} 
\sum_{j\in\mathcal{B}_1}\sum_{k\in\mathcal{B}_2}
J_{jk}^{(\{\sigma\})} 
e^{i(\tilde{\varphi}_{j}-\tilde{\varphi}_{k})},
\end{equation}
 the inter-sector cotunneling Hamiltonian is given by
\begin{equation}\label{eq:intersecproj}
H_{\mathcal{B}_1\mathcal{B}_2} = 
J_{\mathcal{B}_1\mathcal{B}_2}^{\left(\{\sigma\}\right)}\sigma^1\cdots\sigma^n 
e^{i(g_1\Phi_1-g_2\Phi_2)} +\mathrm{h.c.}
\end{equation}
The full strong-coupling tunneling Hamiltonian  follows by summing over all subsector pairs.
Several comments are now in order:
\begin{enumerate}[(i)]
\item
The above discussion also holds if one of the subsectors ${\cal B}_{1,2}$ contains just a single lead,
where Eq.~\eqref{eq:intersecproj} applies as soon as the other subsector enters strong coupling.
\item 
Phase differences between individual $\tilde{\varphi}_{j}$ (or $\tilde{\varphi}_{k}$) in Eq.~\eqref{eq:interseccoupling} are pinned by the potential terms in Eq.~\eqref{eq:reduce}.
Therefore also the inter-sector differences $\tilde{\varphi}_{j}-\tilde{\varphi}_k$ are fixed, and all contributions to  $J_{\mathcal{B}_1\mathcal{B}_2}^{(\{\sigma\})}$ 
in Eq.~\eqref{eq:interseccoupling} add up with a collective inter-sector phase $\varphi_{\mathcal{B}_1\mathcal{B}_2}^{(\{\sigma\})}$.
\item   Equation~\eqref{eq:intersecproj} implies a drastic reduction in the number of boson fields 
 at strong coupling. However, the parameter $g_0$ in Eq.~\eqref{eq:comphase} implies that the
collective fermionic lead obtained from $\phi_0$ in general will represent an interacting fermion theory.
 To see this, we note that $\tilde g=1/g_0^2=M$ acts like a Luttinger liquid 
 parameter \cite{Nayak1999,Beri2017}. For $M>1$, we thus have attractive electron-electron interactions.
We note in passing that RG couplings between isotropized subsectors acquire
the same enhancement factor $\sim M= \tilde g$, see Sec.~\ref{sec3} and Ref.~\cite{Beri2017}.
\item We may encounter multiple tunneling paths with distinct Pauli strings connecting both subsectors,
in particular, for systems with closed loops. 
The strong-coupling Hamiltonian then contains a center-of-mass term as in Eq.~\eqref{eq:intersecproj}
for each of these non-equivalent tunneling paths. Their relative phase,
\begin{equation}
\varphi^\mathrm{loop} = \varphi_{\mathcal{B}_1\mathcal{B}_2}^{(\{\sigma\})} - 
\varphi_{\mathcal{B}_1\mathcal{B}_2}^{(\{\sigma' \})},
\end{equation}
 coincides with the loop phase in Eq.~\eqref{eq:loophase}.
\end{enumerate}

We emphasize that the strong-coupling projection of bosonic subsectors to center-of-mass fields is not limited to a specific setup.
In particular, the same idea allows one to elegantly discuss nonequilibrium effects due to applied bias voltages in simply-coupled systems \cite{Beri2017}, see also App.~\ref{appC}.
For the resulting effective models,
similar to the discussion in Sec.~\ref{sec3a}, our approach only depends on whether tunneling paths between a pair of subsectors contain overall commuting or anticommuting Pauli strings.
For mutually commuting operators, we
arrive at RG equations as in Eqs.~\eqref{eq:subsectorRG} and \eqref{eq:intersubRG}.
Now consider two tunneling operators with couplings 
$J_{\mathcal{B}_1\mathcal{B}_2}^{(\{\sigma\})}$ and $J_{\mathcal{B}_2\mathcal{B}_3}^{( \{ \sigma' \})}$, which connect
subsector ${\cal B}_2$ with subsectors ${\cal B}_1$ and ${\cal B}_3$, 
respectively, cf.~App~\ref{appA}.
If the corresponding Pauli strings anticommute,  no RG contributions will be generated for arbitrary couplings 
 $J_{\mathcal{B}_1\mathcal{B}_3}^{(\{\sigma^{\prime\prime} \})}$ between ${\cal B}_1$ and ${\cal B}_3$.
 However, if two (or more) paths between a \emph{pair} of subsectors contain anticommuting Pauli strings, 
 one obtains the hybridization and feedback contributions discussed in Sec.~\ref{sec3a}.

\subsection{Decoupling fields via hybridization terms}\label{sec4b}

A second key ingredient concerns a decoupling of certain linear
combinations of boson fields from cotunneling operators with Pauli strings.
Such strategies go back to work of Emery and Kivelson (EK) \cite{Emery1992} and are often used for Kondo systems,
see, e.g., Refs.~\cite{Gogolin1998,Fabrizio1995,Landau2017b}. In particular, they show
that the relevant low-energy degrees of freedom at strong coupling usually differ from those at weak coupling.
After an orthogonal rotation of the original set of lead boson fields $\{\phi_j(x)\}$ to a new set 
of boson fields $\{ \phi_\alpha(x) \}$, 
which corresponds to a highly non-local operation in terms of the underlying fermions,
one performs a unitary rotation involving Pauli operators and the boundary phase fields $\Phi_\alpha=\phi_\alpha(0)$.  
One can thereby trade off the coupling of some boson species with a Pauli operator 
in favor of a hybridization term. These generalized EK decoupling schemes can allow for exact results
at special parameter choices (Toulouse points) \cite{Gogolin1998}, where the bare hybridization, cf.~Sec.~\ref{sec3}, is 
precisely compensated by the effects of the unitary transformation. 

\subsubsection{Center-of-mass (charge) field decoupling}

We first discuss this strategy for systems with near-degenerate box charge states
described by a spin operator ${\bf S}_a$ for box $a$, see Eq.~\eqref{eq:Hchargespin} and Sec.~\ref{sec2c}.
This idea was discussed for the single-impurity TKE in Refs.~\cite{Herviou2016,Michaeli2017,Landau2017}. For 
our more general systems with Pauli operators and several boson fields, our approach differs only in the type of 
fields that are decoupled.
While for one near-degenerate box, one can decouple the center-of-mass (`charge') field \cite{Herviou2016,Michaeli2017,Landau2017},
for two (or more) coupled near-degenerate boxes, one should first project to the combined lowest-energy  
charge state. For example, in the notation of Eq.~\eqref{eq:Hchargespin}, we have
\begin{equation}
H_{ab} \simeq \Delta E_a S_z^a+\Delta E_b S_z^b+  \sum_{j,k} \left(
t_{jk}\gamma_j\gamma_k S_{+,a}S_{-,b} +\mathrm{h.c.}\right),
\end{equation}
with MZMs $\gamma_{j/k}$ on box $a/b$, respectively.  Note that the total inter-box tunneling amplitude,
$t_{ab} = \sum_{j,k} t_{jk}\gamma_j\gamma_k$, fluctuates as it depends on the Majorana parities $i\gamma_j\gamma_k = \pm1$. For
 nearly charge-degenerate cases, we have $|t_{ab}|\gg \Delta E_{a/b}\sim \Delta n_{g,a/b}$, and one can project onto
 the subspace spanned by the lowest-energy total charge states, e.g., $\ket{0}_{ab} = \ket{0_a1_b}$ and $\ket{1}_{ab} = \ket{1_a0_b}$ in the notation of Sec.~\ref{sec2c}.

Using this strategy, one arrives at a single (or conglomerate of strongly coupled)  box(es) attached only to leads 
on the outside. With total-charge states described by a collective spin variable $\bf{S}$, one finds a  general tunneling 
Hamiltonian as in Eq.~\eqref{eq:Hchargespin}. Using the center-of-mass field $\Phi_0$ 
and reduced fields $\tilde \Phi_k$ as in Sec.~\ref{sec4a}, we obtain the boundary term,
 see also \cite{Herviou2016,Michaeli2017},
\begin{equation} 
H_{b} = \sum_{j,k}\left(\lambda_{jk}  \gamma_j\kappa_k S_+ e^{ -i (\tilde{\Phi}_k+ \frac{1}{\sqrt{M}}\Phi_0) }+\mathrm{h.c.}\right)+\Delta E S_z,
\label{eq:Hchargespin2}
\end{equation}
where $\Delta E\sim \Delta n_g$ is the overall detuning energy between the total  charge states $S_z = \pm 1/2$. We 
now notice that an EK-type unitary rotation, $U = e^{-i\frac{1}{\sqrt{M}}\Phi_0S_z}$, 
can decouple the center-of-mass field $\Phi_0$.
The tunneling Hamiltonian, $\tilde H_b=U H_b U^\dagger$, is then of the form
\begin{equation}
\tilde{H}_b = \sum_{j,k}\left( 
\lambda_{jk}\gamma_j\kappa_k S_+ e^{-i\tilde{\Phi}_k}+\mathrm{h.c.}\right)+(\Delta E-\Lambda\Theta'_0) S_z, \label{eq:Hchargespin3}
\end{equation}
where the term $\Lambda \Theta'_0 S_z$ comes from the transformation of $H_{\rm leads}$.
Clearly, by tuning $\Delta E\sim\Delta n_g$, one could quench the last (hybridization) term in Eq.~\eqref{eq:Hchargespin3}.  
The reduced field combinations $\tilde{\Phi}_k$ do not change the box charge state anymore due to the constraint $\sum_k\tilde{\Phi}_k = 0$. 
Rather these new fields describe injection of a single electron from lead $k$, which is then transmitted into all outer leads
with the same probability. Together with the isotropization of the $\lambda_{jk}$ couplings, this constitutes a 
hallmark for the TKE \cite{Herviou2016,Michaeli2017}.  We therefore expect TKE physics to be ubiquitous in systems of coupled near-degenerate boxes. 

\subsubsection{Relative (spin) field decoupling}

Following a similar strategy, we now give an example for how to decouple relative (`spin') fields in the cotunneling regime of 
charge-quantized coupled box systems. We focus on the single-MZM two-lead junction  described by the junction Hamiltonian $H_{1,2}$ in 
Eq.~\eqref{eq:Htun12}, see Fig.~\ref{fig2}(c) and Sec.~\ref{sec2c}, where the boson fields $\Phi_{x,y}$ refer to the two leads coupled to a single MZM.
 
We first switch to linear combinations of the lead bosons, $\phi_{c,s}(x)=(\phi_x(x)\pm \phi_y(x))/\sqrt2$, and analogously for the conjugate $\theta_\nu$ fields. 
As shorthand, we will just write $\Phi_{c} = (\Phi_x+\Phi_y)/\sqrt2$ and $\Phi_s=(\Phi_x-\Phi_y)/\sqrt2$, with the implicit understanding that
the transformation is also carried out in the bulk.   
From Eq.~\eqref{eq:Htun12}, we then obtain  
\begin{equation}\label{eq:Htun12rot}
H_{1,2} = \left(\lambda_x\sigma_x e^{i\frac{\Phi_s}{\sqrt2}}+\lambda_y\sigma_y e^{-i\frac{\Phi_s}{\sqrt2}}\right)
e^{i(\frac{\Phi_c}{\sqrt2}-\varphi)}+\mathrm{h.c.},
\end{equation}
where only the $\Phi_s$ field couples in an essential manner to the Pauli operators $\sigma_{x,y}$.

At this point, we apply the unitary transformation $U = e^{i\sigma_z\Phi_s/\sqrt2}$. Switching to 
$\sigma_\pm = (\sigma_x\pm i\sigma_y)/2$, the transformed junction Hamiltonian, $\tilde H_{1,2}=UH_{1,2}U^\dagger$, is given by 
\begin{eqnarray}\nonumber
\tilde H_{1,2}&=& \left[\lambda_x\left(\sigma_+ + \sigma_- e^{\sqrt2 i  \Phi_s}\right)-i\lambda_y\left(\sigma_-+\sigma_+ e^{-\sqrt2 i \Phi_s}\right) \right]
\\ &&\qquad \times \
e^{i(\frac{\Phi_c}{\sqrt2}-\varphi)} + {\rm h.c.}
\end{eqnarray}
In addition, transformation of the lead Hamiltonian generates a hybridization
term $(v_F/\sqrt2)\sigma_z\Theta_s'$.
The $\lambda_{x/y}$ terms now contain rapidly oscillating phase exponentials of $\Phi_s$.
In the spirit of the rotating-wave approximation, we drop such highly irrelevant tunneling operators.  
We then obtain the boundary Hamiltonian 
\begin{equation}\label{eq:EKdecoupling}
\tilde{H}_b =\left(\lambda_x\sigma_+ -i\lambda_y\sigma_- \right)e^{i(\frac{\Phi_c}{\sqrt2}-\varphi)}+\mathrm{h.c.}+
\Lambda\sigma_z\Theta_s',
\end{equation}
where $\Lambda$ includes a bare coupling value and the above $v_F/\sqrt2$ term.
The field $\Phi_s$ has thus been decoupled at the cost of an interaction between the lead density $\sim\Theta_s'$ and the Pauli operator 
$\sigma_z$.  However, at the special Toulouse point, $\Lambda=0$, the spin-field combination  disappears completely. We note that for the example discussed here, an equivalent 
decoupling can also be achieved with a fermionic representation of the leads.

In the  remainder of this section, see also Sec.~\ref{sec5}, we employ the above ideas  
to study the strong-coupling regime for the applications discussed from the weak-coupling RG perspective in Secs.~\ref{sec3b}--\ref{sec3d}. 

\subsection{Two-box device}\label{sec4c}

For the two-box device in Fig.~\ref{fig3}, see Sec.~\ref{sec3b}, according to our strategy in Sec.~\ref{sec4a},
we first identify the important boson fields that should be kept in the strong-coupling analysis. There are four
such fields, namely the center-of-mass fields for the left/right box,
 $\Phi_{L/R}$, with $g_{L/R} = 1/\sqrt{M_{L/R}}$ in Eq.~\eqref{eq:comphase}, 
 and the left/right central lead fields, $\Phi_{l/r}$, with $g_{l/r} = 1$.
We then have five different inter-sector couplings: $J_{Z}, J_{X,l/r}$, and $J_{Y,r/l}$.
Since those effective couplings are obtained by summing over individual leads, they include 
enhancement factors $\sim M_{L,R}$, cf.~Sec.~\ref{sec4a}.
From the cotunneling Hamiltonian in Eqs.~\eqref{eq:HtwoboxL} and \eqref{eq:HtwoboxLR},  
the effective strong-coupling theory follows as 
\begin{equation}\label{eq:2boxeff}
H_\mathrm{eff} = 
\sum_{\nu=L,R,l,r}H_\mathrm{leads}[\phi_\nu,\theta_\nu]-\frac12\left(\Gamma_b+\Gamma_b^\dagger\right), 
\end{equation}
with the boundary operator
\begin{eqnarray}\label{eq:Gtwoboxeff}
\Gamma_b &=& J_{X,l}\sigma_x e^{i\left(\Phi_l-g_L\Phi_L\right)}
+J_{X,r}\sigma_x e^{i\left(\Phi_r-g_R\Phi_R\right)} \\ \nonumber
&+& J_{Y,l}\sigma_y e^{i\left(\Phi_l-g_R\Phi_R\right)}
+ J_{Y,r}\sigma_y e^{i\left(\Phi_r-g_L\Phi_L\right)}\\ \nonumber
&+&iJ_Z\sigma_z e^{i\left(g_L\Phi_L-g_R\Phi_R\right)}.
\end{eqnarray}

For arbitrary device parameters, further analytical progress is difficult even though always at least one of 
the charge/spin combinations of the central lead fields,  
$\Phi_{c,s}=(\Phi_l\pm\Phi_r)/\sqrt2$, can be decoupled by an EK transformation, see Sec.~\ref{sec4b}.
For instance, when studying transport between $L/R$ leads, a decoupling of $\Phi_s$
is most sensible.
In any case, numerical approaches can provide another option to investigate the physics encoded by Eq.~\eqref{eq:Gtwoboxeff}, 
e.g., via quantum Monte Carlo simulations \cite{Egger1998} or the numerical renormalization group \cite{Galpin2014}.

We here instead focus on a simpler yet nontrivial two-box setup
which does allow for analytical progress. Such a device is shown in Fig.~\ref{fig1}, where in contrast to the case
depicted in Fig.~\ref{fig3}, we now only have a single central lead ($\Phi_l$).
The strong-coupling Hamiltonian for this device follows directly from Eqs.~\eqref{eq:2boxeff} and 
\eqref{eq:Gtwoboxeff} by putting $J_{X/Y,r}= 0$.   The remaining couplings are given by 
\begin{equation}
J_{x}= J_{X,l}, \quad J_y=J_{Y,l}, \quad J_z=J_Z.
\end{equation}
We then perform an EK transformation with
$U = e^{i\sigma_z(\Phi_l-g_R\Phi_R)}$.  Following the steps in Sec.~\ref{sec4b}, 
the transformed Hamiltonian, $\tilde H_{\rm eff}=H_{\rm leads}+\tilde H_b$, contains the 
boundary term
\begin{eqnarray}
\tilde H_b &=& -\frac12\left(\tilde\Gamma_b+\tilde\Gamma_b^\dagger\right)
+ \Lambda \sigma_z \left( \Theta_l'-g_R \Theta'_R \right) ,\\ \nonumber 
\tilde\Gamma_b&=&  (J_x \sigma_+ - i J_z \sigma_z)  e^{-i(g_L\Phi_L-g_R\Phi_R)}-iJ_y \sigma_+.
\end{eqnarray}
 The hybridization parameter $\Lambda=\Lambda_0-v_F$ includes
a bare coupling $\Lambda_0$, where $v_F$ is due to the EK transformation of $H_{\rm leads}$.
Next, we perform an orthogonal rotation of the $\phi_{L/R}(x)$ phase fields,
\begin{equation}\label{eq:orthrot}
\left( \begin{array}{c} \phi_1\\ \phi_2\end{array}\right)
=  \frac{1}{\bar g}
\left( \begin{array}{cc} g_L & -g_R \\  g_R & g_L 
\end{array}\right)
\left( \begin{array}{c} \phi_L\\ \phi_R\end{array}\right), \quad \bar g=\sqrt{g_L^2+g_R^2},
\end{equation}
resulting in
\begin{eqnarray} 
\tilde H_b &=& - \frac12\left( (J_x \sigma_+ - iJ_z \sigma_z) e^{-i\bar g\Phi_1}
+{\rm h.c.} + J_y\sigma_y\right)  \nonumber \\ &+&\frac{\Lambda}{\bar g}\sigma_z \left(
\bar g\Theta_l'+g^2_R\Theta_1'+g_Rg_L\Theta_2'\right).
\label{finalh}
\end{eqnarray}
The setup with $M_L=M_R=2$ in Fig.~\ref{fig1} now gives access to an exact solution 
at the Toulouse point, $\Lambda=0$, via the refermionization approach \cite{Gogolin1998}.
Indeed, for $\bar g=1$, which only holds for $M_{L}=M_R=2$, the operator $e^{-i\bar g\Phi_1}$ 
in Eq.~\eqref{finalh} can be expressed as fermion annihilation operator (up to a Klein factor),
and $\tilde H_{\rm eff}$ thus reduces to a noninteracting fermion theory for $\Lambda=0$.
 In the remainder of this subsection, we thus assume $M_L=M_R=2$ as in Fig.~\ref{fig1}, 
 but for now still allow for $\Lambda\ne 0$. 

At this stage, we employ Eq.~\eqref{eq:boso} backwards to obtain chiral fermion operators $\psi_\nu(x)$
associated with the respective boson field $\phi_\nu$ with mode index $\nu=1,2,l$.
Using $\Psi_\nu = \psi_\nu(0)$ and  recalling that $\Psi^\dagger_\nu\sim\kappa_\nu e^{i\Phi_\nu}$, see
Eq.~\eqref{eq:boso}, Klein factors ($\kappa_\nu$) are again represented as Majorana operators.
 In addition, we express Pauli operators as Majorana bilinears, 
$\sigma_{\alpha=x,y,z}=i\gamma_\alpha\gamma_0$, with the overall parity constraint
$\gamma_0\gamma_x\gamma_y\gamma_z=1$.  
We now notice  (i) that $\kappa_{\nu=1}$ is the only Klein factor which explicitly appears in $\tilde H_{\rm eff}$, 
and (ii) that $i\gamma_0\kappa_1=\pm 1$ is conserved.
Choosing $i\gamma_0\kappa_1=-1$, Eq.~\eqref{finalh} yields 
\begin{eqnarray}\nonumber 
&&\tilde H_b=J_x  \gamma_x 
\left(\Psi_1^\dagger-\Psi_1\right) + i \left( J_x \gamma_y-J_z\gamma_z \right) 
\left(\Psi_1^\dagger+\Psi_1\right) \nonumber\\ \label{finalHa}
&&- \ \frac{iJ_{y}}{2} \gamma_z\gamma_x+ i \Lambda \gamma_y\gamma_x
:2\Psi_l^\dagger\Psi_l^{} +
\Psi^\dagger_1\Psi_1^{} - \Psi^\dagger_2\Psi_2^{} :,
\end{eqnarray}
where $:\ :$ indicates normal-ordering and $1/\sqrt{\alpha}$ factors from the short-distance 
cutoff in Eq.~\eqref{eq:boso} have been absorbed in $J_{x,z}$.
Clearly, in the Toulouse limit, we indeed have noninteracting fermions.
In the final step, we switch to chiral Majorana fermions by writing 
\begin{equation}\label{eq:referm}
\psi_\nu(x)=\left[\xi_{\nu}(x) +i\eta_{\nu}(x)\right]/\sqrt2,
\end{equation}  
where $\xi_\nu(x)=\xi^\dagger_\nu(x)$ and $\eta_\nu(x)=\eta^\dagger_\nu(x)$ 
obey the algebra $\{ \xi_\nu(x),\eta_{\nu'}(x')\}=\delta(x-x')\delta_{\nu\nu'}$ and so on \cite{Gogolin1998}.
The bulk Hamiltonian then takes the form
\begin{equation}
H_{\rm leads}=-\frac{i v_F}{2}\sum_{\nu}\int_{-\infty}^\infty dx\left(\xi_\nu\partial_x\xi_\nu+\eta_\nu\partial_x\eta_\nu\right),
\end{equation}
and the Toulouse Hamiltonian is given by
 \begin{eqnarray} \label{finalh5}
 H_{\rm Toul} &=& H_{\rm leads}-i 
 \sqrt2  J_x \gamma_x \eta_1(0) \\ 
 & +& i\sqrt2 (J_x \gamma_y-J_z\gamma_z)\xi_1(0) -\frac{iJ_y}{2} \gamma_z\gamma_x .
 \nonumber
 \end{eqnarray}
Interaction corrections come from the $\Lambda$-term in Eq.~\eqref{finalHa},  
\begin{equation}\label{eq:hlm}
H_\Lambda= i\sum_{\nu=1,2,l}\Lambda_\nu \gamma_y\gamma_x\xi_\nu(0)\eta_\nu(0),
 \end{equation}
with couplings $\Lambda_\nu\sim\Lambda$.
The corrections are RG irrelevant. In fact,
for $J_{x,y,z}\neq 0$, they have scaling dimension $d_{\nu=l,2}=3$ and $d_{\nu=1}=2$, respectively.
Finally, noting that $\Psi_1\sim e^{-i\Phi_1}= e^{-i(\Phi_L-\Phi_R)/\sqrt2}$, we observe that the central lead ($\Psi_l$) 
decouples at the Toulouse point, i.e., no current will flow through this lead.
A detailed discussion of nonequilibrium transport for this setup is given in Sec.~\ref{sec5}.

\subsection{Single MZM coupled to multiple leads}\label{sec4d}

Our next example is that of a single MZM coupled to two or three leads, see Sec.~\ref{sec3c}.
Recall that this case derives from the two-box setting by taking $M=M_L$ leads connected by simple lead-MZM contacts, while $M_R=1$ for the
right box (boson field $\Phi_z$). In addition, we have two central leads ($\Phi_{x,y}$).
With the effectively isotropic Hamiltonian in Eq.~\eqref{eq:HtunB}, the construction of the strong-coupling theory 
then proceeds precisely as in Sec.~\ref{sec4c}. 
In fact,  $H_{\rm eff}$ follows directly by setting $M_R = 1$ in Eq.~\eqref{eq:Gtwoboxeff}. 
Using the center-of-mass field $\Phi_L = g_L\sum_{j=1}^M\Phi_j$ with $g_L=1/\sqrt{M}$, 
Eq.~\eqref{eq:2boxeff} holds with
\begin{equation}
\Gamma_b =\sum_{\alpha=x,y,z} J_\alpha\sigma_\alpha e^{i\left(g_L\Phi_L-\Phi_\alpha\right)},
\label{eq:GtunB}
\end{equation}
where the couplings $J_{\alpha}$ have been specified in Eq.~\eqref{fourcot}. 

This strong-coupling Hamiltonian again represents an interacting problem.
However, for $J_z  = 0$, analytical progress can be made by using the charge/spin fields $\Phi_{c,s}$
instead of $\Phi_{x,y}$. As discussed in Sec.~\ref{sec4b}, 
 the EK transformation $U=e^{i\sigma_z\Phi_s/\sqrt2}$ decouples $\Phi_s$ 
  from $\Gamma_b$ and generates a hybridization term from $H_{\rm leads}$.
Moreover, by an orthogonal rotation $(\phi_L, \phi_c)\to (\phi_a,\phi_0)$, cf.~Eq.~\eqref{eq:orthrot},
we switch to the linear combinations 
\begin{eqnarray}\label{eq:phia}
 \phi_a &=& \frac{1}{g_a}\left( g_L\phi_L - \frac{1}{\sqrt2}\phi_c\right),\\  \nonumber
 \phi_0 & =& \frac{1}{\sqrt{M+2}}\left(\phi_x+\phi_y+\sum_{j=1}^M\phi_j\right),
\end{eqnarray}
with the parameter
\begin{equation}\label{eq:gadef} 
g_a=\sqrt{g_L^2+1/2}=\sqrt{\frac{M+2}{2M}}.
\end{equation}
The field $\phi_0$ is nothing but the total center-of-mass phase field for all $M+2$ leads, which decouples 
from the transport problem. We hence obtain
\begin{eqnarray}
\tilde{H}_{\rm eff} &=& \sum_{\nu= a,s}
H_\mathrm{leads}[\phi_\nu,\theta_\nu] -\frac12\left(\tilde{\Gamma}^{}_b+\tilde{\Gamma}_b^\dagger\right) + \Lambda_s\sigma_z\Theta_s',
\nonumber \\
\tilde{\Gamma}_b &=&\left(J_x\sigma_+ + iJ_y\sigma_-\right) e^{ig_a\Phi_a}.\label{eq:H12strong}
\end{eqnarray}
This Hamiltonian describes collective charge transport between the $M$ outer leads 
and the charge field $\Phi_c=(\Phi_x+\Phi_y)/\sqrt2$, where 
the Pauli operators $\sigma_{x,y}$ couple only to $\Phi_a$, cf.~Eq.~\eqref{eq:phia}.
We find $\Lambda_s=\Lambda_0-v_F/\sqrt2$ with the bare hybridization $\Lambda_0$.

In general, this is an interacting theory even at the Toulouse point, $\Lambda_s = 0$. Indeed, refermionization
of the $\phi_a$ channel implies attractive electron-electron interactions since
 $\tilde g_a=1/g^2_a>1$ for $M>2$, see Eq.~\eqref{eq:gadef}.
The only exception to this rule arises for $M=2$, where $g_a=1$ and refermionization obtains a
 noninteracting fermion theory for $\Lambda_s=0$.  We thus put $M=2$ and
refermionize the two remaining lead channels $\nu=a,s$ as in Sec.~\ref{sec4c}.  
In addition, we again write Pauli operators as bilinears of Majorana operators, 
$\sigma_{\alpha=x,y,z}=i\gamma_\alpha\gamma_0$, with 
$\gamma_0\gamma_x\gamma_y\gamma_z=1$. Using the  fermion operator
 $d=\left(\gamma_x+i\gamma_y\right)/2$, we thus have 
\begin{equation}
\sigma_+=\sigma_-^\dagger=id\gamma_0, \quad \sigma_z=1-2d^{\dagger}d,
\end{equation}
and the tunneling operator $\tilde \Gamma_b$  in Eq.~\eqref{eq:H12strong} has the form
\begin{equation}
\tilde \Gamma_b=i\gamma_0\kappa_a  \left( J_x d+i J_yd^{\dagger}\right) \Psi_a^{\dagger},
\end{equation}
where the cutoff in Eq.~\eqref{eq:boso} has been absorbed in $J_{x,y}$.
Clearly, the local parity $i\gamma_0\kappa_a$ is conserved. Choosing $i\gamma_0\kappa_a=+1$,
we get the boundary contribution to $\tilde H_{\rm eff}=H_{\rm leads}+\tilde H_b$ in the form
\begin{eqnarray}\nonumber 
 \tilde H_b  &=&   -\frac12 J_x\left(\Psi_a^{\dagger}d+d^{\dagger}\Psi_a\right)-\frac{i}{2} 
 J_y\left(\Psi_a^{\dagger}d^{\dagger}-d\Psi_a\right)\\ &-& 
 \Lambda_s\left(2d^{\dagger}d-1\right) :\Psi_s^{\dagger}\Psi^{}_s:
\end{eqnarray}
Using $J_{\pm}=(J_y\pm J_x)/2\sqrt2$ and the chiral Majorana fermion representation in Eq.~\eqref{eq:referm}, 
we can alternatively write  
\begin{equation}\label{eq:hhh}
\tilde H_b = iJ_+\xi_a(0)\gamma_x+iJ_-\eta_a(0)\gamma_y + \Lambda_s\gamma_x\gamma_y\xi_s(0)\eta_s(0).
\end{equation}

Remarkably, the just obtained effective strong-coupling Hamiltonian $\tilde H_{\rm eff}$ for the setup in Fig.~\ref{fig1} 
coincides with the asymmetric two-channel Kondo model studied in detail in Ref.~\cite{Fabrizio1995}.    
Let us briefly summarize the corresponding physics.
First, in the channel-symmetric case, $J_-=0$, the system shows non-Fermi liquid behavior 
at the Toulouse point, $\Lambda_s=0$. The leading irrelevant operator $\sim\Lambda_s$ 
has scaling dimension $d=3/2$ which determines the power-law exponent of the 
temperature- and/or voltage-dependent conductance \cite{Gogolin1998}. 
 For $J_-\ne 0$, on the other hand, the Toulouse Hamiltonian 
obtained from Eq.~\eqref{eq:hhh} is a sum of two independent Majorana resonant level models
and thus exhibits Fermi liquid behavior at low energy scales.  Furthermore,
at the Toulouse point but otherwise for arbitrary $J_\pm$, exact results for the full counting statistics
of nonequilibrium transport have been derived by Gogolin and Komnik \cite{Gogolin2006}. 
Their results immediately apply to the present setting, see also Sec.~\ref{sec5}.

\subsection{Loop qubit}\label{sec4e}

Last we turn to the strong-coupling regime of the loop qubit device depicted in 
Fig.~\ref{fig4}. While a limiting case of the problem, cf.~Eq.~\eqref{eq:Hqppstrong} below, 
has already been addressed in Ref.~\cite{Plugge2016b}, in view of
the present experimental interest in this device, we here give a more complete picture. 
Following the strategy in Sec.~\ref{sec4a}, we first define a center-of-mass field for the $M$ outer leads, 
$\Phi_L = g_L\sum_{j=1}^M\Phi_j$ with $g_L=1/\sqrt{M}$.
We also recall that $\Phi_c$ denotes the boson field for the central lead contacting two MZMs on the box, see Fig.~\ref{fig4}.
Our weak-coupling analysis in Sec.~\ref{sec3d} has then identified two qualitatively different candidate strong-coupling fixed points. 

The first type is stable and describes an RG flow towards loop phase $\varphi_0 =\pm\pi/2$.
Without loss of generality, we choose $\varphi_0 = +\pi/2$, where one has a 
strong complex-valued cotunneling amplitude $L_+$ and a vanishing amplitude $L_-$ in Eq.~\eqref{eq:Hqpp}. 
We then obtain the strong-coupling theory, $H_{\rm eff}=H_{\rm leads}+H_{\varphi_0=\pi/2}$, with
\begin{equation}
H_{\varphi_0=\pi/2} = -J_+\sigma_+ e^{i(g_L\Phi_L - \Phi_c)} + {\rm h.c.}
+\Lambda\sigma_z\Theta_c',
\label{eq:Hqppstrong}
\end{equation}
where $J_+=ML_+/\sqrt2$ and $\Lambda=2(\Lambda_c-\tilde\Lambda)$, see Sec.~\ref{sec3d}.
For  $M = 1$, Ref.~\cite{Plugge2016b} found 
that this model can be mapped onto a fully anisotropic single-channel Kondo model.
For $M\geq 2$, as we discuss below, the central lead $\Phi_c$ instead dynamically decouples from the 
outer leads which in turn develop a TKE for $M\ge 3$.  

The second fixed point, taken as $\varphi_0 = 0$ without loss of generality, is unstable with respect to phase variations $\delta\varphi_0$, see Sec.~\ref{sec3d}. This fixed point is qualitatively different from the first one, as it implies $L_+=L_-$ 
and $\Lambda\sim\sin\varphi_0=0$.
The strong-coupling theory follows from Eqs.~\eqref{eq:HtunC} and \eqref{eq:Hqpp},
\begin{equation}
H_{\varphi_0=0} = -(J_x\sigma_x+J_y\sigma_y) e^{i(g_L\Phi_L - \Phi_c)} + {\rm h.c.}\label{eq:Hqppstrong2}
\end{equation}
with $J_{x,y} \sim\lambda_{x,y}$ in Eq.~\eqref{eq:HtunC}. Next we use the local fermion parity representation
of Pauli operators, $\sigma_{x,y} = i\gamma_{x,y}\kappa$. 
Since both $J_{x}$ and $J_y$ are real, with fixed ratio during the RG flow, we can construct a new Majorana operator 
\begin{equation}
\gamma = (J_x\gamma_x+J_y\gamma_y)/J,\quad J=\sqrt{J_x^2+J_y^2}.
\end{equation}  
The central contact thus couples to a single Majorana operator $\gamma$ only, since 
the relative tunneling phase between the lead and the two original MZMs is zero (or $\pi$).  
For other values of $\varphi_0$, such a reduction is not possible.
However,  the above reasoning is not restricted to  the 
cotunneling regime. The same steps also apply for the tunneling Hamiltonian in Eq.~\eqref{eq:HtunC},
and hence we expect this effect to always appear so long as $\varphi_0= 0$~mod~$\pi$.  Finally, 
we note that Eq.~\eqref{eq:Hqppstrong2} has conserved fermion parity $i\gamma\kappa = \pm 1$. Choosing 
 $i\gamma\kappa=1$, we obtain
\begin{equation}\label{eq:beri}
H_{\varphi_0=0} = - 2J \cos(g_L\Phi_L - \Phi_c).
\end{equation}
Using the results of Ref.~\cite{Beri2017}, where Eq.~\eqref{eq:beri} also appears, 
we thus have access to the full nonequilibrium transport characteristics 
between the central lead and an arbitrary number $M\geq 2$ of outer leads.

The loop qubit device in Fig.~\ref{fig4} is likely most relevant as a starting point to more 
complicated Majorana multi-junctions and networks.  To guide such experimental tests, let us briefly 
summarize how quantum transport is expected to depend on the loop phase $\varphi_0$.
First, since experiments are performed at small but finite temperature and bias, features of the unstable fixed
point should appear in a region around $\varphi_0 = 0~\mathrm{mod}~\pi$ with small but non-zero hybridization.
Now consider the case $M=1$.
If $\varphi_0 \approx 0$, our theory predicts qualitatively the same behavior as for a two-terminal mesoscopic Majorana wire  \cite{Fu2010,Zazunov2011,Hutzen2012}. While transport for half-integer $n_g$, i.e., at a charge-degeneracy point, exhibits 
the quantized zero-temperature conductance $G_0 = e^2/h$,  transport in the cotunneling regime will be strongly
suppressed.  Conversely, as one increases $\varphi_0$, the conductance should approach $G_0$ largely independent of $n_g$. Tunneling of charges then is not 
due to charge-degenerate states but rather caused by a Kondo resonance \cite{Plugge2016b}. 
The latter arises due to many-body screening of the spin-$1/2$ impurity $\sim (\sigma_x,\sigma_y,\sigma_z)$ built from three Majorana operators, 
two at the central and one at the simply-coupled lead.
Next we consider the case $M\geq 2$, i.e., a multi-terminal measurement of conductance between the central lead and outer leads in Fig.~\ref{fig4}.
Starting again with $\varphi_0\approx 0$, the device should display the transport behavior expected for the TKE 
\cite{Beri2012,Altland2013,Beri2013,Zazunov2014,Altland2014,Beri2017}, with fractional conductance values at zero 
temperature and non-Fermi liquid power laws in the temperature- and/or voltage-dependent conductance. 
In the loop qubit device, a natural experiment includes probing the finite-bias conductance through the 
central lead, which for $\varphi_0\approx 0$ should reveal the features discussed in Ref.~\cite{Beri2017}.
For increasing $\varphi_0$, the ensuing hybridization $\Lambda$ at the central (and all other) leads 
will gap out the Majorana fermion pair involved in $\sigma_z =i\gamma_y\gamma_x$.
As a consequence, transport involving the central lead will be blocked 
at  temperatures and/or voltages below the Kondo temperature $T_K$ of the box.
We thus predict drastically different low-energy conductance behavior depending on both the loop phase
$\varphi_0$ and on the number of attached leads.

Finally, tuning the system to near half-integer $n_g$ is not expected to qualitatively affect the
above conclusions for $M\ge 2$, cf.~Secs.~\ref{sec2c} and \ref{sec4b}. 
However, the Kondo temperature is expected to strongly depend on $n_g$ 
\cite{Herviou2016,Michaeli2017}.
Therefore, while the approach to a universal conductance value in the strong-coupling regime takes place 
independent of the loop phase $\varphi_0\neq 0$ and of the gate parameter $n_g$,
the finite-energy behavior will depend on those parameters.

\section{Transport in a two-box device}
\label{sec5}

In this section, we study nonequilibrium transport properties for the two-box device in Fig.~\ref{fig1}
by employing the strong-coupling theory  in Sec.~\ref{sec4c}. 
We consider the system right at the Toulouse point, with the
noninteracting Hamiltonian $H_{\rm Toul}$ in Eq.~\eqref{finalh5}.
The resulting physics is expected to be generic since interaction corrections around the
Toulouse point, see Eq.~\eqref{eq:hlm}, are RG irrelevant. 
For closely related models, an exact solution for the full counting statistics of charge transport 
has been described in Refs.~\cite{Gogolin2006,Landau2017b}. In what follows, we adapt those results 
to  the setup in Fig.~\ref{fig1}.

To that end, we first recall that at the Toulouse point, the central lead $\psi_l$ will dynamically decouple from the
transport problem, see Sec.~\ref{sec4c}. However, a small residual current is expected to flow through the central lead due to RG irrelevant interaction corrections not considered below.
We thus focus on a transport configuration, where the $M_L=2$ ($M_R=2$) leads 
attached via simple contacts to the left (right) box are held at chemical potential $+eV/2$ ($-eV/2$).  
In particular, there are no applied voltages between leads attached to the same box.
If the latter were present, quick equilibration of leads at each box is expected due to the large intra-sector coupling. 
In contrast, the inter-box coupling may be small and equilibration is perturbed by the central non-simple junction.
We then consider the outcome of a two-terminal measurement of the fluctuating time-dependent current, $I(t)$, 
flowing between individual pairs of leads on different sides.
(The relation to collective inter-sector transport is discussed below and in App.~\ref{appC}.)
During a measurement time $t_m$, 
the charge $q=\int_0^{t_m} dt' I(t')/e$ is transferred between the two leads, where 
the full counting statistics of $q$ follows from a cumulant generating function $\chi(\lambda)$.
In particular, by taking derivatives with respect to the counting field $\lambda$,  
one obtains all cumulants from the relation $\langle\delta^n q\rangle=(-i)^n\partial^n_{\lambda}\ln\chi(\lambda= 0)$.
Below we only discuss the average current, $I$, and the current noise, $S$, which are given by
\begin{equation}\label{curshot}
I= \frac{e}{t_m}\left\langle \delta q\right\rangle,\quad S=\frac{2e^2}{t_m}\left \langle \delta^2 q\right\rangle.
\end{equation}
We next relate transport between individual leads attached to the left and right box, respectively, to the transformed fermion
basis at strong coupling, cf.~Sec.~\ref{sec4c}.
To this end, observe that application of the operator $\Psi_1\sim e^{-i(\Phi_L-\Phi_R)/\sqrt2}$ on 
an arbitrary system state amounts to transporting one unit of charge between the left and right side.
Recalling the center-of-mass phases
 $\Phi_L = (\Phi_{L_1}+\Phi_{L_2})\sqrt{2}$ and 
$\Phi_R = (\Phi_{R_1}+\Phi_{R_2})\sqrt{2}$  in
 terms of the physical leads $L_{1,2}$ and $R_{1,2}$, per tunneling event, 
the charge transferred at each individual lead hence is $e^*=e/2$.
One thus can include the counting field by letting 
$\Psi_1\to e^{+(-)i\lambda/4}\Psi_1$ on the forward (backward) time branch of 
the Keldysh partition function for $H_{\rm Toul}$ \cite{Gogolin2006}.
Since the projected theory in Eq.~\eqref{finalh5} contains only $\Psi_1$, the inclusion of a counting field is relevant only for
 one out of the four fermion species in the ensuing two-channel Kondo model \cite{Landau2017b}.
 
After some algebra along the steps in Refs.~\cite{Gogolin2006,Landau2017b}, where only the 
Green's functions for the three impurity-Majorana operators $\gamma_{x,y,z}$ in Eq.~\eqref{finalh5} 
have to be updated, we obtain the zero-temperature generating function,
\begin{equation}\label{FiniteBias}
 \ln\chi(\lambda)=\frac{t_m}{2\pi}\int_0^{eV/2} d\omega\ln\left(1+{\cal T}(\omega) [e^{i\lambda}-1 ] \right),
\end{equation}
with the frequency-dependent transparency
\begin{equation}\label{Transparency0}
{\cal T}(\omega)=\frac{\left(\Gamma_z \omega^2 -  \Gamma_x J_y^2\right)^2}
{( \Gamma_x^2+\omega^2)\left[(\omega^2-J_y^2)^2+
\omega^2( \Gamma_x+\Gamma_z)^2\right]}.
\end{equation}
We here define the energy scales $\Gamma_{x,z}\sim J_{x,z}^2$, where the proportionality constant also takes into account 
the rescaling of $J_{x,z}$ due to the short-distance cutoff in Eq.~\eqref{eq:boso}, see Sec.~\ref{sec4c}.
We mention in passing that the finite-temperature variant of Eq.~\eqref{FiniteBias} can readily be 
expressed in terms of Eq.~\eqref{Transparency0} as well, cf.~Refs.~\cite{Gogolin2006,Landau2017b}.

Let us then discuss the predictions of Eq.~\eqref{FiniteBias} for the current-voltage characteristics
and for shot noise in this system.

\subsection{No Majorana hybridization: $J_y=0$}\label{sec5a}

We start with the case $J_{y}=0$, where the MZM operators $\gamma_{x}$ and $\gamma_z$ are not hybridized.
Defining the channel hybridizations
\begin{equation}\label{channelhyb}
\Gamma_1=\Gamma_x,\quad \Gamma_2=\Gamma_x+\Gamma_z,
\end{equation} 
Eq.~\eqref{Transparency0} takes the simpler form 
\begin{equation}\label{Transparency1}
 {\cal T}_{J_y=0}(\omega)=\frac{(\Gamma_1-\Gamma_2)^2\omega^2}{(\omega^2+\Gamma_1^2)  (\omega^2+\Gamma_2^2)}.
\end{equation}
Equation \eqref{Transparency1} gives the transparency of two competing Majorana channels  coupled
by the respective channel hybridization $\Gamma_{1,2}$ to a single impurity, and therefore describes
the asymmetric two-channel Kondo effect \cite{Fabrizio1995,Landau2017b}.
In fact, after a rotation of the impurity-Majorana sector, $H_{\rm Toul}$ in Eq.~\eqref{finalh5} directly corresponds to the 
Hamiltonian in Eq.~\eqref{eq:hhh}, with $\Gamma_{1/2}=\Gamma_{-/+}\sim J^2_{-/+}$.
The current-voltage characteristics readily follows from  Eqs.~\eqref{curshot}--\eqref{Transparency1},
\begin{equation}\label{eq:nopar}
 I =\frac{e}{h}\frac{\Gamma_2-\Gamma_1}{\Gamma_2+\Gamma_1} \left[ \Gamma_2
 \tan^{-1}\left(\frac{eV}{2\Gamma_2}\right) - \Gamma_1  \tan^{-1}\left(\frac{eV}{2\Gamma_1}\right) \right] .
\end{equation}
It is instructive to consider several limiting cases of Eq.~\eqref{eq:nopar}. 

First,  the current \eqref{eq:nopar} between the left and the right side vanishes identically for the channel-symmetric case
with $\Gamma_2-\Gamma_1=\Gamma_z\to 0$.
In fact, this result makes sense because the dependence of $\Gamma_z$ on the microscopic tunnel amplitudes 
implies that both boxes are decoupled in that limit, 
$\sqrt{\Gamma_z}\sim J_z \sim \lambda_L\lambda_R t_{LR}/E_C^2\to 0$.

Second, a related observation is that by increasing $\Gamma_x$ at a fixed value of $\Gamma_z$, 
the current in Eq.~\eqref{eq:nopar} will also decrease. 
Indeed, for $\Gamma_x/\Gamma_z\to\infty$,  Eq.~\eqref{channelhyb} implies that  
we effectively come back to the limit $\Gamma_1=\Gamma_2$ again, where the current vanishes.
We note that in order to increase $\sqrt{\Gamma_x}\sim J_x\sim\lambda_L\lambda_l/E_C$ at fixed $\Gamma_z$, the tunnel coupling $\lambda_l$ between the left box and the central lead has to increase.
Although charge transfer at the central contact is dynamically blocked, the coupling $\Gamma_x$ still has profound effects on the 
system. In particular, for $\Gamma_x\ne 0$, the central junction is effectively driven out of resonance by a misalignment 
of the spin direction $\sim (\sigma_x,\sigma_y,\sigma_z)$ with respect to the left-right transport direction $\sim \Gamma_z$.

Finally, in  the opposite limit $\Gamma_x/\Gamma_z\to 0$, we instead approach the single-channel case 
with transparency
\begin{equation}\label{Transp1}
{\cal T}_{\Gamma_x=J_y=0}(\omega)= \frac{\Gamma_z^2}{\omega^2+\Gamma_z^2},
\end{equation} 
where we note that $\Gamma_1=\Gamma_x= 0$ in Eq.~\eqref{channelhyb} implies $\Gamma_2=\Gamma_z$.
From Eq.~\eqref{curshot}, we obtain for $(eV,\Gamma_x) \ll \Gamma_z$ the transport observables
\begin{eqnarray} \label{eq:aa}
 I&=&\frac{e}{2h}\left[eV-2\Gamma_x\tan^{-1}\left(\frac{eV}{2\Gamma_x}\right)\right],\\ \nonumber
 S&=&\frac{2e^2}{h}\left[\frac{\Gamma_x}{2}\tan^{-1}\left(\frac{eV}{2\Gamma_x}\right)-\frac{\Gamma_x^2}{\left(eV\right)^2+4\Gamma_x^2}eV\right].
\end{eqnarray}
Defining the backscattered current $I_b=(e^2/2h)V-I$, we see that for $\Gamma_x\ll eV\ll \Gamma_z$, the shot noise power is given by $S=2e^{*}I_b$ with elementary charge $e^{*}=e/2$.
The shot noise comes from the weakly coupled ($\Gamma_1$) channel, while the strongly coupled ($\Gamma_2$) channel is fully transmitted (with the two-channel Kondo value of the conductance, $G=e^2/2h$) and thus noiseless. 
Equation \eqref{eq:aa} yields the same fractional Fano factor, $F=S/2I_b=e^*/e=1/2$, as recently found in a related two-channel charge Kondo system \cite{Landau2017b}.
In our case, a single additional Majorana operator enters the low-energy theory for $\Gamma_x>0$, given by the Klein factor $\kappa_{l}$ at the central lead, see Fig.~\ref{fig1}.
In the Toulouse-point Hamiltonian $H_\mathrm{Toul}$ in Eq.~\eqref{finalh5} it is represented by the Majorana operator $\gamma_x$.
This causes the backscattering processes in Eq.~\eqref{eq:aa}, described by the fractional charge $e^*=e/2$.

\begin{figure}[t]
\centering  
\includegraphics[width=0.45\textwidth]{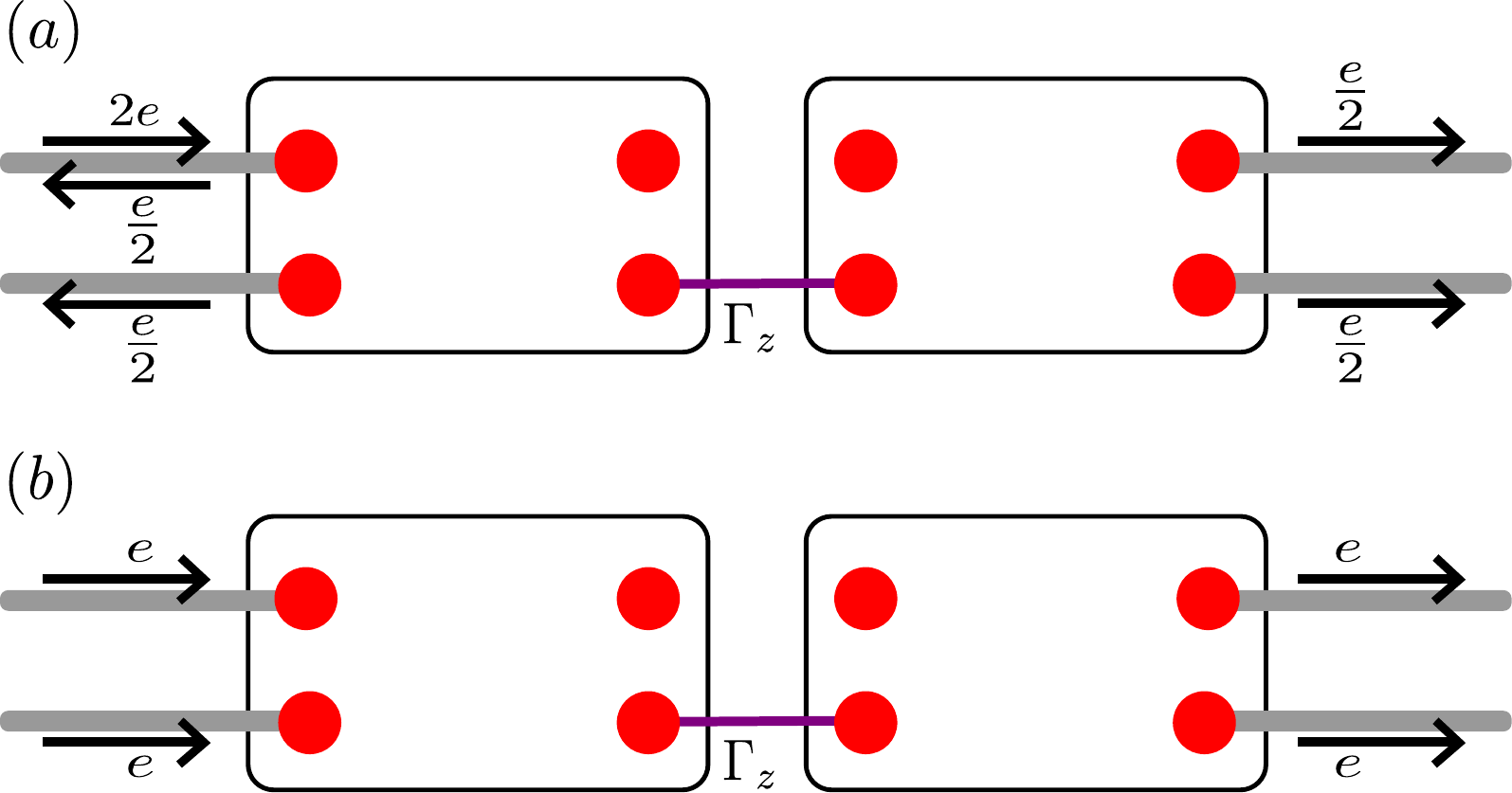}
\caption{
Cross-correlated Andreev reflections (AR) generated from individual correlated AR processes in the two-box device with $J_y=0$ and $\Gamma_x\to 0$, see Sec.~\ref{sec5a}.
(a) A single AR at the top left lead, followed by the emission of charge $e/2$ into all four leads, 
forms a correlated AR process as in the TKE~\cite{Altland2013,Beri2013,Zazunov2014}. 
Since formation of charge dipoles between the left leads is suppressed by the strong intra-sector coupling, a nonequilibrium excitation is left behind.
(b) A sequence of two correlated ARs, one each at top and bottom left leads, comprises a cross-correlated AR. This allows for the 
cotunneling of a Cooper pair by subsequent crossed ARs between left (in) and right (out) leads.
Further discussion, see main text and App.~\ref{appC}.}
\label{fig6}
\end{figure}

For $\Gamma_x\to 0$, we also can draw an interesting link to the single-impurity TKE.
Indeed, since the left and right boxes are now joined by a strong coupling $\Gamma_z$,
this two-box setup should be related to the TKE for 
a single large box with $M=M_L+M_R=4$ attached leads, cf.~Sec.~\ref{sec3b}.
Taking into account results by B{\'e}ri \cite{Beri2017},
we offer a detailed discussion of this correspondence in App.~\ref{appC}.
The subsector-biased case considered here, with applied voltage $V_{L,R}=\pm V/2$ for 
all leads with $j\in\mathcal{B}_L$ and $k\in\mathcal{B}_R$, respectively, is slightly more involved than the one in Ref.~\cite{Beri2017}.
For the two-terminal conductance measurement in Eq.~\eqref{eq:aa},
we here find $G_{jk} = e^2/2h$ between any pair of individual leads $j$ and $k$. Instead,
 for collective inter-sector transport, we show in App.~\ref{appC} that the left-right conductance is given by 
 $G_{LR} = 2e^2/h$. The latter arises by summing the current over all leads in the respective subsectors, 
 and it comprises cross-correlated Andreev reflections involving the Cooper pair charge $e^*_{LR} = 2e$.
The generation of these processes is detailed in Fig.~\ref{fig6} and App.~\ref{appC}.
We thus predict the appearance of different effective charges due to hybridization with the central lead ($e^* = e/2$) and 
due to finite-energy corrections in collective left-right inter-sector transport ($e^*_{LR} = 2e$).

\subsection{Finite Majorana hybridization}\label{sec5b}

Next we include the effects of a finite Majorana hybridization $J_y\ne 0$. 
In order to obtain a qualitative understanding, we first analyze the limit $J_y\gg {\rm max}( \Gamma_{x,z},eV)$, where 
the impurity term $-i(J_y/2) \gamma_z\gamma_x$ in $H_{\rm Toul}$
implies the fixed parity $i\gamma_z\gamma_x=+1$. 
Equation \eqref{finalh5} can therefore be projected to a simpler single-channel model,
$H'_{\rm Toul}=H_{\rm leads}  + i\sqrt2 J_x \gamma_y\xi_1(0)$, where a single  MZM  ($\gamma_y$)
is coupled to a single chiral Majorana mode ($\xi_1$). The parity constraint $i\gamma_z\gamma_x=+1$ 
here effectively blocks the other chiral Majorana channel $\sim\eta_1$.
Indeed, for $J_y\to \infty$,  the general transparency expression in Eq.~\eqref{Transparency0} reduces to
the single-channel result
\begin{equation}\label{Transpstronghyb}
{\cal T}_{J_y\to\infty}(\omega)= \frac{\Gamma_x^2}{\omega^2+\Gamma_x^2},
\end{equation} 
but with active channel $\sim\Gamma_x$ instead of $\Gamma_z$ in Eq.~\eqref{Transp1}. We thus come back to single-channel
results for conductance and shot noise again, with $\Gamma_x$ as the only remaining parameter.
Left-right transport then takes place exclusively by cotunneling via the central lead $l$ in Fig.~\ref{fig1}.

\begin{figure}[t]
\centering  
\includegraphics[width=0.45\textwidth]{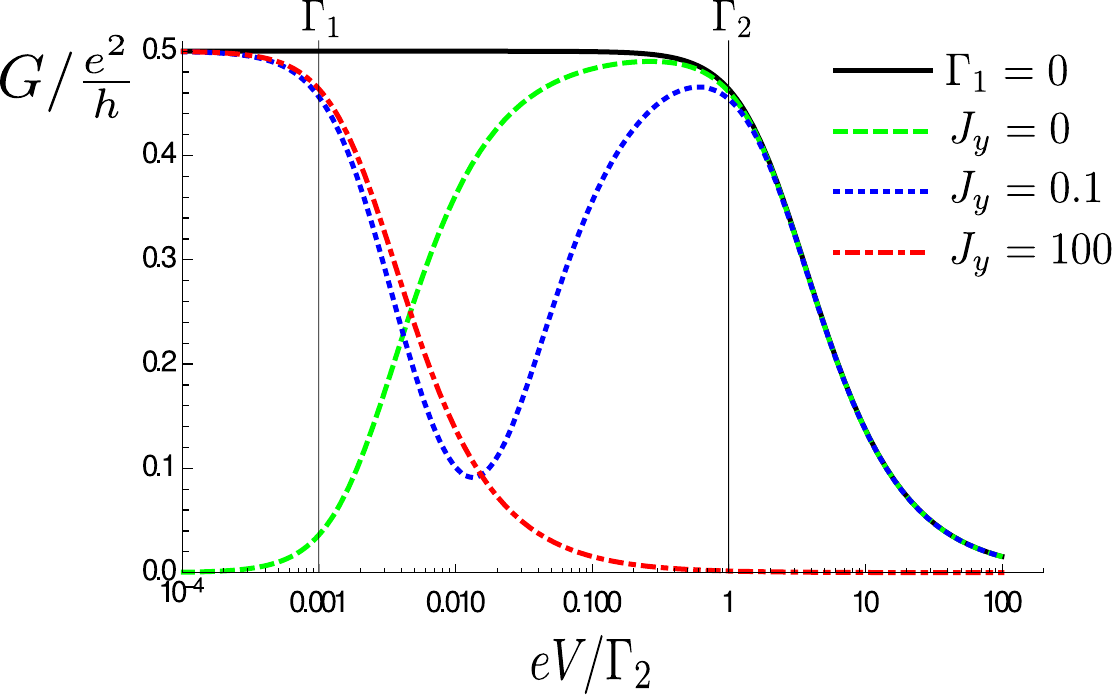}
\caption{
Two-terminal conductance $G=I/V$ vs voltage $V$ between two leads attached to different boxes in the two-box device of Fig.~\ref{fig1}.
The shown results hold at the Toulouse point, see Eq.~\eqref{finalh5}, and follow from Eqs.~\eqref{curshot}--\eqref{Transparency0}. For detailed discussion, see main text.}
\label{fig7}
\end{figure}

We next discuss the voltage dependence of the nonlinear conductance $G=I/V$, which is plotted for typical parameters 
in Fig.~\ref{fig7}. The shown curves have been obtained by 
numerical evaluation of Eqs.~\eqref{curshot}--\eqref{Transparency0}.  
First, the conductance for $\Gamma_1=J_y=0$ (black solid curve) illustrates
the single-channel case in Sec.~\ref{sec5a}, where Eq.~\eqref{eq:aa} gives $G=e^2/2h$ for 
$eV\ll \Gamma_2$, in accordance with Fig.~\ref{fig7}.
Second, turning to $\Gamma_1\ll \Gamma_2$ but still keeping $J_y=0$ (dashed green curve, with $\Gamma_1/\Gamma_2=0.001$), we observe that 
the conductance vanishes at very low voltages but recovers to a large value near $e^2/2h$ 
within the window $\Gamma_1\ll eV\ll \Gamma_2$.  Such a behavior is consistent with our analytical result in 
Eq.~\eqref{eq:nopar}, which describes the asymmetric two-channel Kondo effect with 
two competing Majorana channels coupled to an impurity.

The remaining two curves in Fig.~\ref{fig7} include the effects of a finite Majorana hybridization $J_y$, which now can cause antiresonances or resonances in the voltage dependence of the conductance.
First, for $J_y\gg {\rm max}( \Gamma_{1,2}, eV)$, cf.~the red dash-dotted curve for $\Gamma_1/\Gamma_2=0.001$
and $J_y/\Gamma_2=100$, two of the three impurity-Majorana operators $\gamma_{x,y,z}$
are gapped out by the large $J_y$. We thus observe single-channel physics of the weaker channel, with 
coupling $\Gamma_1 = \Gamma_x$ in Eq.~\eqref{Transpstronghyb}.
Next, for $\Gamma_1\ll J_y\ll \Gamma_2$ (blue dotted curve, $\Gamma_1/\Gamma_2=0.001$ and $J_y/\Gamma_2=0.1$),
after approaching the single-channel value at $eV\simeq \Gamma_2$, the voltage dependence of the conductance reveals an antiresonance for $\Gamma_1\lesssim eV\lesssim J_y$ with subsequent recovery at $eV\lesssim \Gamma_1$. Here, in the low-bias regime, a combined channel as in Eq.~\eqref{Transpstronghyb} is activated.
Finally, for general non-zero couplings $\Gamma_{1,2}$ and $J_y$, we observe a complex interplay between the asymmetric two-channel Kondo effect and impurity hybridization phenomena.
However, for our case with three coupled impurity-Majorana operators, 
the low-frequency transparency in Eq.~\eqref{Transparency0} always approaches the unitary limit, $\mathcal{T}(\omega\to 0) = 1$.
This behavior can be rationalized by noting that at sufficiently low energies, one (rotated) Majorana pair will effectively
be gapped out for $J_y\neq 0$. The remaining third Majorana operator then remains free.
This MZM provides a single-channel transport resonance pinned to the Fermi level, with the universal zero-bias conductance $G=e^2/2h$.

We conclude that the device in Fig.~\ref{fig1} allows for a complete solution of the nonequilibrium transport
problem at the Toulouse point.
An interesting open question for future research will be to address interaction corrections around this point,
which can easily be included in the full counting statistics formalism used above \cite{Gogolin2006,Landau2017b}.

\section{Concluding remarks}
\label{sec6}

In this work, we have studied quantum transport through coupled Majorana box devices.  
Since Majorana boxes represent an attractive platform for realizing topological qubits,
coupled box devices are of present interest for quantum information processing applications, see, e.g., 
Refs.~\cite{Plugge2017,Karzig2017}.  When normal leads are tunnel-coupled to such a system, the spin-$1/2$ 
degrees of freedom representing Majorana box qubits will be subject to Kondo screening via cotunneling
processes, culminating in the topological Kondo effect \cite{Beri2012}.  Consequently, when different
boxes are connected, one encounters competing Kondo effects and related phenomena in a non-Fermi 
liquid setting.

For general systems of this type, we have introduced a powerful and versatile theoretical framework for studying the low-energy physics and quantum transport. Our theory employs Abelian bosonization of the lead fermions together
with the Majorana-Klein fusion method of Refs.~\cite{Altland2013,Beri2013}.  
For a single box, the resulting problem
is purely bosonic and admits an asymptotically exact solution for the corresponding non-Fermi liquid fixed point \cite{Altland2013,Beri2013}. 
However, for coupled-box systems, we found that
additional local sets of Pauli operators due to non-conserved local fermion parities must be taken into account. 
Despite the complexity of the resulting problem, it is possible to make analytical progress.   
Approaching the physics both from the weak-coupling side (see our RG analysis in Sec.~\ref{sec3}) and 
in the strong-coupling regime (see our effective low-energy theory for 
the most relevant collective degrees of freedom in Sec.~\ref{sec4}), a rich interplay between different types
of single- or multi-box topological Kondo effects has been encountered.  

We have in detail examined the transport characteristics of the three perhaps most basic devices  
where non-conserved fermion parities play a central role.  One of these includes the loop qubit device suggested in Ref.~\cite{Karzig2017}.    
Importantly, the methods put forward in this work also allow one to obtain nonperturbative transport 
results in moderately complex setups.  This aspect should be especially valuable in view of the fact that transport measurements could give clear and unambiguous nonlocality signatures for Majorana states in such devices.
At the fundamental level, non-simple lead-MZM junctions can not be described by purely 1D non-branched networks that admit a solution in terms of the 
Majorana-Klein fusion approach, cf.~Sec.~\ref{sec2c}.
Therefore transport measurements in our setups may reveal more 
profound signatures of Majorana non-Abelian statistics when compared to the 
simple junction setups considered in experiments so far.
While a detailed discussion of alternative non-Majorana transport 
scenarios, e.g., for the loop qubit device in Fig.~\ref{fig4}, 
is beyond the scope of our work, we hope that our predictions will 
soon be put to an experimental test.

\begin{figure}
\centering  
\includegraphics[width=0.45\textwidth]{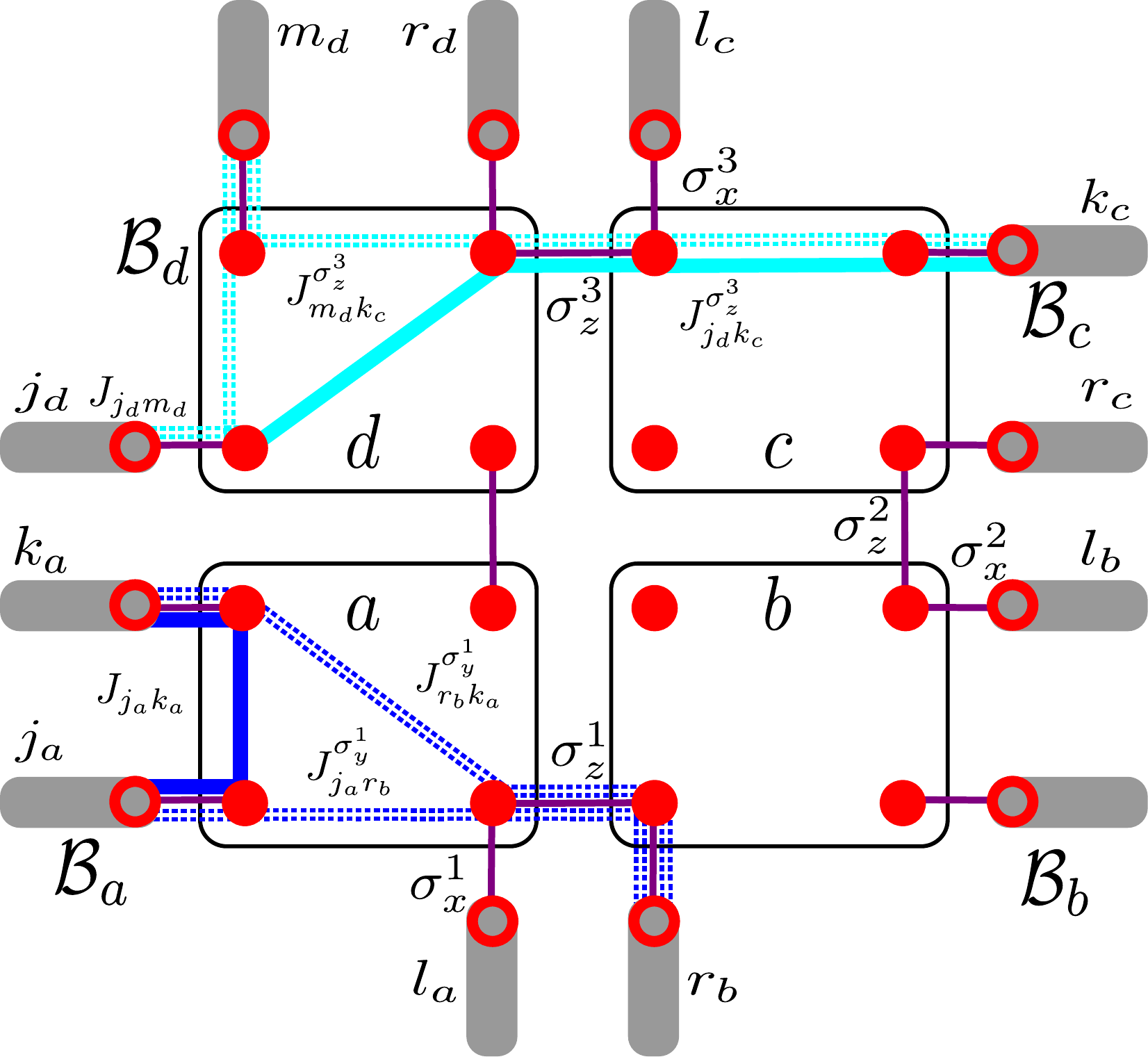}
\caption{Example for a coupled Majorana box device with four boxes ($a,b,c,d$), with symbols as in Figs.~\ref{fig1}
and \ref{fig2}.
The bosonic subsectors ${\cal B}_{a,b,c,d}$ contain $M_a=M_d=2$ and $M_b=M_c=1$ 
leads with simple lead-MZM contacts  to the respective box. The device has four MZM-MZM tunnel bridges and
three pairs of central leads [$(l_a,r_b)$, $(l_b,r_c)$, and $(l_c,r_d)$]
with non-simple lead-MZM contacts.  Each central lead also forms its own subsector. Non-conserved local fermion 
parities are encoded by Pauli operators $\sigma_{x,y,z}^{m=1,2,3}$.	
We also illustrate how RG terms arise from contractions of cotunneling operators:
(i) For $j_a\ne k_a\in {\cal B}_a$, the second term in Eq.~\eqref{eq:subsectorRG} is due to
contraction of $J_{j_a r_b}^{(\sigma_y^1)}$ and $J_{r_b k_a}^{(\sigma_y^1)}$ (dashed dark blue line)
which renormalizes $J_{j_ak_a}$  (solid dark blue).
(ii)
For lead indices $j_d\ne m_d\in{\cal B}_d$, the contraction of $J_{j_dm_d}$ and $J_{m_dk_c}^{(\sigma_z^3)}$ (dashed cyan)  
renormalizes the amplitude $J_{j_dk_c}^{(\sigma_z^3)}$ (solid cyan), cf.~Eq.~\eqref{eq:intersubRG}.}
\label{figA1}
\end{figure}

\acknowledgements
We thank A. Altland, F. Buccheri, K. Flensberg, L.A. Landau, C. Mora, E. Sela, and A. Zazunov for discussions. This work has been supported by the Deutsche Forschungsgemeinschaft within CRC TR 183 (project C01) and within Grant No.~EG 96/11-1.

\appendix
\section{Examples for RG contributions}\label{appA}

\begin{figure}
\centering  
\includegraphics[width=0.45\textwidth]{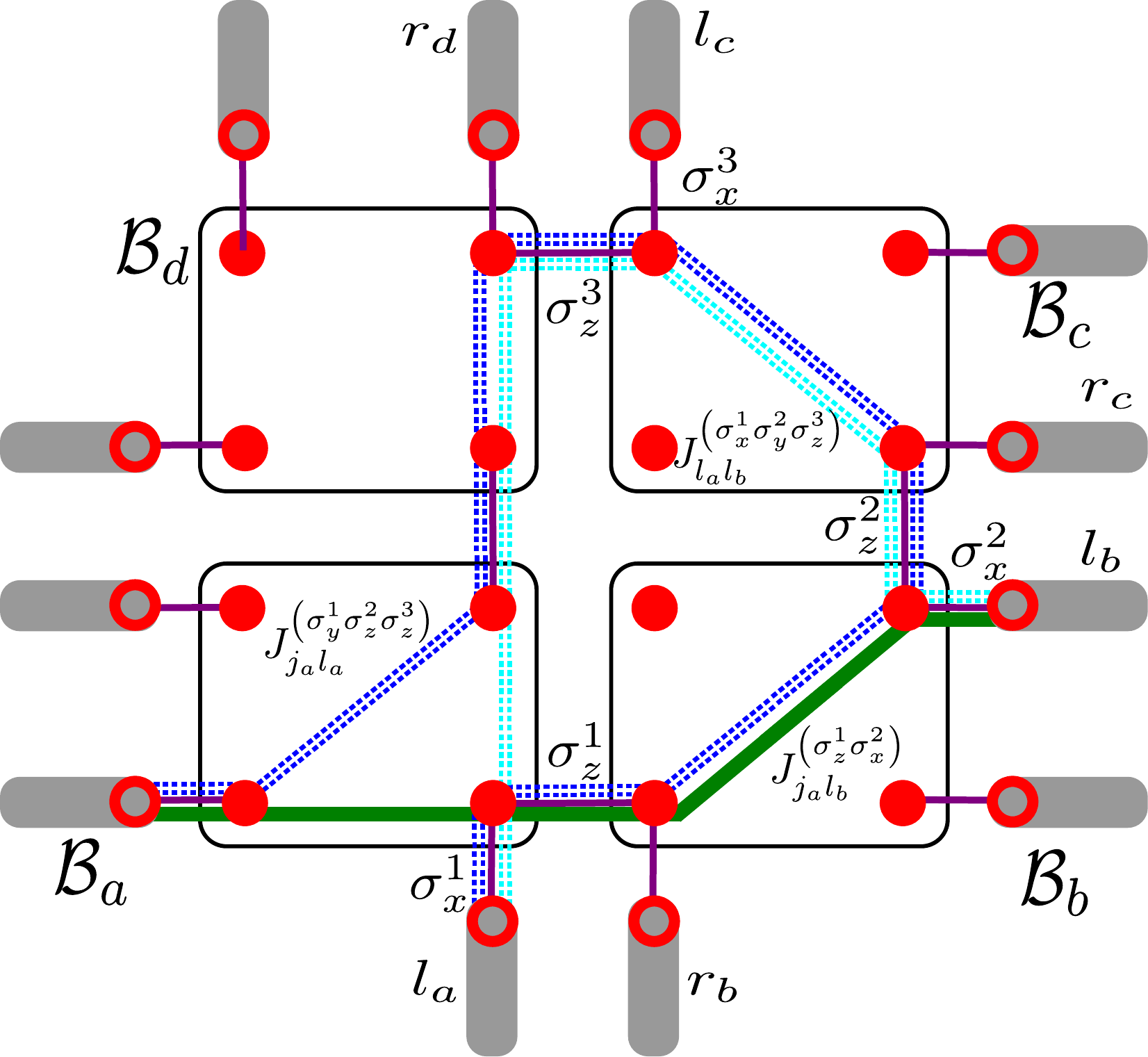}
\caption{Illustration of additional contributions to the RG flow of cotunneling amplitudes 
connecting lead $j_a\in {\cal B}_a$ and lead $l_b$ beyond those specified in Eq.~\eqref{eq:intersubRG},
using the same the device as in Fig.~\ref{figA1}. 
By contracting the two cotunneling operators with amplitudes $J_{j_a l_a}^{\left(\sigma_y^1\sigma_z^2\sigma_z^3\right)}$ (dashed dark blue) and $J_{l_a l_b}^{\left(\sigma_x^1\sigma_y^2\sigma_z^3\right)}$ (dashed cyan), the composite 
Pauli string is given by $\left(\sigma_y^1\sigma_z^2\sigma_z^3\right)\left(\sigma_x^1\sigma_y^2\sigma_z^3\right)\sim
\sigma_z^1\sigma_x^2$. This contraction contributes to the RG flow of $J_{j_a l_b}^{\left(\sigma_z^1\sigma_x^2\right)}$ (green solid).}
\label{figA2}
\end{figure}

We here give further details and examples for the general RG equations in Sec.~\ref{sec3a}, which we illustrate for 
a device with four coupled Majorana boxes, see Fig.~\ref{figA1}.  
We start with two examples for tunneling operators connecting leads in different subsectors and therefore involving Pauli strings.
Our first example, with Pauli string length $n=1$, comes from lowest-order tunneling events connecting 
a  lead $j_a\in{\cal B}_a$ to lead $l_a$ (resp.~$r_b$) in Fig.~\ref{figA1}, where
 the Pauli operator $\sigma_x^1$ (resp.~$\sigma_y^1$) appears in Eq.~\eqref{eq:Hbososub2}.
 Note that lead $l_a$ (resp.~$r_b$) forms its own bosonic subsector, see Sec.~\ref{sec3a}.
As second example, again with $n=1$,
we could pick a tunneling path connecting some lead 
$j_a\in \mathcal{B}_a$ with a lead $k_b\in \mathcal{B}_b$ in Fig.~\ref{figA1}.  In that case,
the Pauli operator $\sigma_z^1$ appears in Eq.~\eqref{eq:Hbososub2}.

Next, we discuss the cotunneling amplitudes $J_{jk}^{(\{\sigma\})}$ appearing 
in Eq.~\eqref{eq:subsectorRG}. Such amplitudes connect
a lead  $j=j_d\in{\cal B}_d$ in a bosonic subsector ${\cal B}_d$
to another lead $k=k_c\notin {\cal B}_d$ which is not part of this subsector,  cf.~Eq.~\eqref{eq:Hbososub2}. 
Here, lead $k$ could be part of the bosonic subsector ${\cal B}_c$ in Fig.~\ref{figA1}.
For  example, taking short tunneling paths connecting leads $j_d$ and $k_c$ in Fig.~\ref{figA1} (cyan lines), 
the Pauli string reduces to $\sigma^3_z$.
Alternatively, lead $k$ may correspond to a non-simple lead-MZM contact. In Fig.~\ref{figA1}, such leads are 
referred to as central leads.  Such a lead forms a bosonic subsector ${\cal B}$ with $M=|{\cal B}|=1$ by itself.
For example, identifying lead $k=r_d$ (resp., $k=l_c$) in Fig.~\ref{figA1}, 
the Pauli string reduces to the single Pauli operator $\sigma_x^3$ (resp., $\sigma_y^3$).
In either case,  pairs of cotunneling
operators will only contribute to the RG flow of $J_{jk}^{(\{\sigma\})}$
if their contraction yields precisely the Pauli string $\sigma^1\cdots\sigma^n$, see Fig.~\ref{figA1} (cyan lines).

The terms on the r.h.s.~of Eq.~\eqref{eq:intersubRG} describe the renormalization of inter-sector cotunneling amplitudes with $j\in{\cal B}_1$ and $k\in{\cal B}_2$
 due to combination of an inter-sector tunneling with intra-sector transitions in either sector $\mathcal{B}_{1,2}$.
On top of this, one can have additional terms that involve intermediate excursions into different sectors $\mathcal{B} \neq \mathcal{B}_{1,2}$.
Such terms have  the schematic form
\begin{equation}\label{eq:A1}
\frac{dJ_{jk}^{(\{\sigma \} ) }}{d\ell}
\sim \sum_{m\notin(\mathcal{B}_1,\mathcal{B}_2)}
J_{jm}^{(\{\sigma^\prime\})} 
J_{mk}^{(\{\sigma^{\prime\prime}\})},
\end{equation}
which contribute only if the contraction of both Pauli strings is consistent with
$(\sigma^{1'}\cdots \sigma^{n'})(\sigma^{1''}\cdots \sigma^{n''}) \sim \sigma^1\cdots\sigma^n$. 
An example for such a process is shown in Fig.~\ref{figA2} using the same system as 
in Fig.~\ref{figA1}. The contracted Pauli strings here share two overlapping anticommuting Pauli operators, and hence overall are commuting.

\begin{figure}[t]
\centering  
\includegraphics[width=0.45\textwidth]{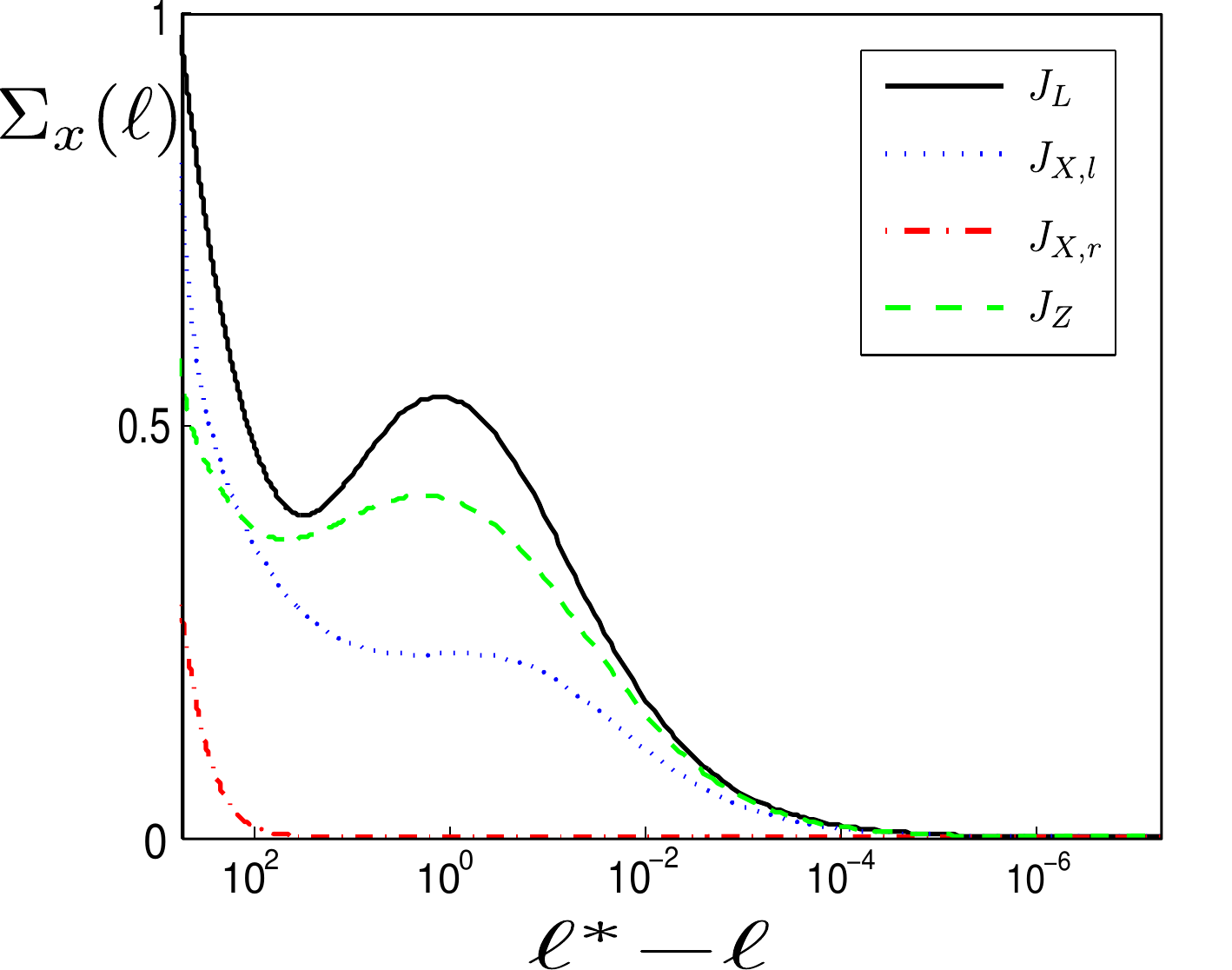}
\caption{RG flow of the anisotropy measures $\Sigma_x$, cf.~Eq.~\eqref{eq:appB2}, for different coupling families $x$  
in the two-box device of Fig.~\ref{fig3}. The weak-coupling RG approach breaks down at $\ell=\ell^*$, where couplings start to
diverge.  We show $\Sigma_x$ vs $\ell^*-\ell$ on a logarithmic scale.  All coupling families become isotropic during the RG flow. }
\label{figA3}
\end{figure}

\section{RG flow for the two-box example}\label{appB}

We here discuss the isotropization of equivalent couplings for the two-box device with $M_L=3$ and $M_R=2$ 
in Fig.~\ref{fig3}, see Sec.~\ref{sec3b}, where equivalence is meant with respect to the
Pauli operator content. In order to check whether the system exhibits isotropization, we perform a numerical integration of 
the RG equations and test how anisotropies present in the bare (initial) couplings develop during the RG flow, 
cf.~Ref.~\cite{Plugge2016b}. Using the couplings in Eqs.~\eqref{eq:HtwoboxL} and \eqref{eq:HtwoboxLR}, we define average couplings
\begin{eqnarray}\nonumber
J_L&=&\frac{1}{M_L(M_L-1)}\sum_{j\neq k\in{\cal B}_L}(J_L)_{jk}, \\ \nonumber
J_{X,l}&=&  \frac{1}{M_L}\sum_{k\in{\cal B}_L}(J_X)_{l k},\\ \label{eq:appB1}
J_{Y,r}&=&\frac{1}{M_L}\sum_{k\in{\cal B}_L} (J_Y)_{rk}, \\ \nonumber
J_Z&=&\frac{1}{M_LM_R}\sum_{j\in{\cal B}_L,k\in{\cal B}_R}(J_Z)_{jk},
\end{eqnarray}
and similarly for $J_R$, $J_{X,r}$ and $J_{Y,l}$.
We then monitor the anisotropy measures, $\Sigma_x(\ell)$, for all seven coupling
 families (indexed by $x$), see Sec.~\ref{sec3b}. These measures are defined from the standard deviation of the 
 coupling family normalized by the respective average value in Eq.~\eqref{eq:appB1}, see also \cite{Plugge2016b},
\begin{equation}\label{eq:appB2}
\Sigma_{J_L}^2=\frac{1}{M_L(M_L-1)}\sum_{j,k\in{\cal B}_L,j\ne k}\frac{\left[(J_L)_{jk}-J_L\right]^2}{J_L^2},
\end{equation}
and likewise for the other coupling families.
Figure \ref{figA3} shows the results of a numerical solution of the RG equations \eqref{eq:RGtwobox1}--\eqref{eq:RGtwobox4} with 
a random choice for the initial couplings, cf.~Ref.~\cite{Plugge2016b}. We have checked that the qualitative behavior
seen in Fig.~\ref{figA3} is largely insensitive to the chosen random realization. 
Fig.~\ref{figA3} shows that all anisotropies become gradually suppressed during the RG flow, which implies effectively isotropic behavior 
within each coupling family and thereby justifies Eq.~\eqref{eq:RGtwoboxiso}.

\section{Biased leads in simply-coupled Majorana boxes}\label{appC}

We here relate our results for the biased two-box setting in Sec.~\ref{sec5} 
with those of B{\'e}ri~\cite{Beri2017}, see also Fig.~\ref{fig6}.
 We first note that for a decoupled central lead in Fig.~\ref{fig1}, in equilibrium we should 
 recover a single-impurity TKE of the combined island with $M=M_L+M_R$ leads. 
 The distinction into different boxes then becomes obsolete. 
 Since Pauli strings are not involved anymore, 
there is no \emph{a priori} reason for a specific partitioning of leads into subsectors. 
However, such a splitting follows from the applied bias voltages in a transport measurement,
 where leads in two subsectors $\mathcal{B}_{a,b}$ are biased relative to each other. In
Sec.~\ref{sec5}, we have considered the case $M_{a,b} = M_{L,R} = 2$, while B{\'e}ri~\cite{Beri2017} investigated
the case of just one biased lead ($M_a=1$) in an otherwise equilibrium $M$-terminal TKE,  $M_b = M-1$.
We next recall the strong-coupling Hamiltonian for this system, see Sec.~\ref{sec4a},
\begin{equation}\label{eq:Hcombiased}
H_{ab} = -J\cos(g_a\Phi_a-g_b\Phi_b) = -J\cos(g\Phi),
\end{equation}
with the collective inter-sector coupling  $J$ and the center-of-mass phase fields 
$\Phi_{a,b}$, cf.~Eq.~\eqref{eq:comphase}, for leads in subsectors $\mathcal{B}_{a,b}$, where 
$g_{a,b}=1/\sqrt{M_{a,b}}$. Equation \eqref{eq:Hcombiased} defines the linear combination $\Phi$ 
with $g=\sqrt{g_a^2+g_b^2}$.

We can now obtain exact nonequilibrium results for charge transport between $\mathcal{B}_{a,b}$ by following the
steps in Ref.~\cite{Beri2017}.  To arrive at a backscattering model from Eq.~\eqref{eq:Hcombiased}, 
one first expresses $\Phi = (\Phi_L+\Phi_R)/\sqrt{2}$ in terms of left- and right-moving chiral boson fields $\phi_{L/R}$. One can then define 
the backscattering interaction $g_\mathrm{bs} =g^2/2$ \cite{Beri2017}, where
Eq.~\eqref{eq:Hcombiased} gives $H_{ab}=-J  \cos[\sqrt{g_{\rm bs}}(\Phi_L+\Phi_R)]$.
The fractional charge $e^\ast$ governing elementary charge transfer processes between subsectors in this 
non-Fermi liquid system is given by the ratio \cite{Beri2017} 
\begin{equation}\label{eq:effcharge}
\frac{e^\ast}{e} = \frac{1}{g_{\rm bs}} = \frac{2M_aM_b}{M_a+M_b}.
\end{equation}
In particular, for $M_a=1$ and $M_b=M-1$, Eq.~\eqref{eq:effcharge} yields  the TKE result for a single biased lead,
$e^\ast_{\mathrm{TKE}} = 2e(M-1)/M$, see Refs.~\cite{Zazunov2014,Beri2017}.
For the symmetric case $M_a = M_b = M/2$, Eq.~\eqref{eq:effcharge} instead gives $e^\ast = eM/2$. 
For instance, putting $M=2$, we confirm that transport is due to cotunneling of electrons~\cite{Fu2010,Zazunov2011,Hutzen2012}.
In our two-box setup with $M=4$, Eq.~\eqref{eq:effcharge} instead gives $e^*_{LR}=2e$.
Transport between the left and right side is thus mediated by the cross-correlated Andreev reflection  (AR)
of Cooper pairs, cf.~Fig.~\ref{fig6}, where one expects the conductance $G_{LR} = 2e^2/h$.
However, in Sec.~\ref{sec5a}, we found that a two-terminal conductance measurement between 
a pair of \emph{individual} leads $j\in\mathcal{B}_L$ and $k\in\mathcal{B}_R$ will give the  
two-channel Kondo value $G_{jk} = e^2/2h$.  The conductance $G_{LR}$ instead follows by summing over 
all participating leads, $G_{LR} = \sum_{j,k} G_{jk} = 2e^2/h$, representing a collective inter-sector conductance 
measurement.

As illustrated in Fig.~\ref{fig6}, one can further reconcile the physics encoded by $e^\ast$
in Eq.~\eqref{eq:effcharge} with previous work on the TKE \cite{Beri2013,Altland2013,Zazunov2014,Beri2017}.
A correlated AR process comprises an AR at one lead (absorbing charge $2e$) along with 
the equal-probability emission of charge $2e/M$ into all $M$ leads, without net charge accumulation on the
island. For a single biased lead, this yields $e^\ast_\mathrm{TKE}$ above.
Next we note that between leads in a biased subsector $\mathcal{B}_a$, charge dipoles 
are forbidden by strong intra-sector couplings. In order to return to an allowed configuration, a 
total of $M_a$ correlated AR events (one from each lead in $\mathcal{B}_a$) have to participate in transport.
Counting after this sequence, each lead in $\mathcal{B}_a$ has emitted charge
\begin{equation}\label{eq:chargea}
q_a = 2e\left[\frac{M-1}{M}-(M_a-1)\frac{1}{M}\right] = 2e\frac{M_b}{M},
\end{equation}
with $M-M_a = M_b$. Similarly, we have $q_b = -2eM_a/M$ absorbed charges per lead in 
$\mathcal{B}_b$, due to $M_a$ split Cooper pairs.
The total, collective charge transported by an effective low-energy process between the two subsectors then is $e^\ast = M_a|q_a|= M_b|q_b|$, as reported in Eq.~\eqref{eq:effcharge}.

From the viewpoint of two-terminal transport between individual leads $j\in\mathcal{B}_a$ and $k\in\mathcal{B}_b$, cf.~Sec.~\ref{sec5}, 
the total outgoing (incoming) charge is democratically distributed into (gathered from) all leads in the opposite sector.
Therefore only the effective charge $e^*_{jk} = q_a/M_b = -q_b/M_a = 2e/M$  is transferred directly from lead $j$ to $k$.
Again summing over leads in the subsectors, one recovers $e^* = \sum_{j,k}e^*_{jk}$.
For our $M=4$ case at hand, in two-terminal transport we reproduce the two-channel Kondo result in Sec.~\ref{sec5a}, $e^\ast_{jk} = e/2$, 
while collective inter-sector transport involves Cooper pairs with $e^\ast_{LR} = 2e$ in Eq.~\eqref{eq:effcharge}.

\end{document}